\documentclass[fleqn,usenatbib]{mnras}

\usepackage{newtxtext,newtxmath}
\usepackage{multirow}
\usepackage{gensymb}

\usepackage[T1]{fontenc}

\DeclareRobustCommand{\VAN}[3]{#2}
\let\VANthebibliography\thebibliography
\def\thebibliography{\DeclareRobustCommand{\VAN}[3]{##3}\VANthebibliography}

\usepackage{graphicx}	
\usepackage{amsmath}	

\title[Be stars in the time domain I]{A study of Be stars in the time domain \\ \large I. Spectral data and polarimetry}

\author[A. Casta\~n\'on Esteban et al.]{
A. Casta\~n\'on Esteban,$^{1}$\thanks{E-mail: a.castanonesteban@2019.ljmu.ac.uk}
I.A. Steele,$^{1}$
H. Jermak$^{1}$
\\
$^{1}$Astrophysics Research Institute, Liverpool John Moores University, 146 Brownlow Hill, Liverpool L3 5RF, UK\\
}

\date{Accepted XXX. Received YYY; in original form ZZZ}

\pubyear{2023}

\begin{document}
\label{firstpage}
\pagerange{\pageref{firstpage}--\pageref{lastpage}}
\maketitle

\begin{abstract}
We present the first part of a spectroscopic and polarimetric study on a sample of 58 Be stars that have been measured since 1998. The aim of the study is to understand the timescales of disk variability, formation and dissipation as a function of the properties (mass, luminosity and rotational velocity) of the underlying B star.  In this paper we classified the sample based on the presence of emission or absorption of the H$\alpha$ line, and the shape of the peak as single or double peak, as well as noting changes between emission and non-emission states.  We find a probability of $\sim 0.75$ percent per year that an object in the sample will undergo such a change.  We also present re-derived values of the projected rotational velocities for the sample. When we compare our polarization values with those from the literature, we find that most of the stars do not show a change in the value of the polarization angle, however a small number show significant changes which could be attributed to either disk strength (optical depth) or geometry changes.  Finally we show how, by combining the (interstellar corrected) degree of polarization and the projected rotational velocity, we can construct an inclination angle-free parameter that includes the true equatorial velocity.  Using this inclination angle-independent parameter we show that the populations of single and double peak stars are indistinguishable, giving further evidence that Be star line profiles are essentially inclination angle driven.

\end{abstract}




\section{Introduction}

Be stars are, on a fundamental level, B stars that show, or have shown in the past,  one or more Balmer lines in emission \citep{collins1987}. The origin of this emission is currently understood as a circumstellar disk around the star that has been created by material lost by the star \citep{poeckert1978,quirrenbach1994}.

The mechanism that creates this disk is still debated: It is understood that a rapid rotational velocity ($v$) plays an important role \citep{slettebak1979}, although in general they do not rotate at the full break-up velocities \citep{porter1996}.  Mechanisms proposed include single-star radial pulsation \citep {baade1988, rivinius1998b}, but it is unclear if this is enough to create the disk \citep{owocki2006, cranmer2009}. Magnetic fields may also play a role \citep{cassinelli2002, brown2008}, but no magnetic fields have been found on Be stars, in contrast with B stars \citep{Wade2014}. Another possibility is that the creation of the disk is due to interaction with a companion star \citep{Kriz1975, Pols1991}, where the accreting star in the transfer not only increases its mass but also its angular momentum \citep{Hastings2021}. A study from \citet{Oudmaijer2010} found that only $\sim$30\% of stars are found in binaries, while a study from \citet{Bodensteiner2020} found that while none of the Be stars studied were in visible binaries with main sequence stars, a large fraction of the fast-rotating B stars were, supporting the idea that binarity may be the cause of the fast rotation and the companion is barely visible anymore after the mass transfer. At the same time, while no main sequence stars have been found as companions of Be systems, subdwarf O and B stars have been found, with characteristics similar to remnant core of a star that did lose most of its envelope \citep{Mourard2015}. A more detailed discussion on the importance of the binarity channel in the creation of the disk around the B star is given by \citet{Jones2022}.

In order to help constrain models, observational studies of the properties and variability of the Be phenomenon are likely to be key.  For example, understanding the timescales of disk loss and reformation as a property of stellar mass, temperature and equatorial velocity is likely to be important \citep{paper5}.  Understanding these dependencies is, however, complicated by the effect of inclination. The disk will have a certain inclination, $i$, with respect to the observation plane and the Earth. As a result, the observed rotational velocity is $v \sin i$. This value is useful as a lower bound for the rotational velocity, but at the same time the actual rotational velocity $v$ can be much higher. An upper bound on the $v$ will be the critical velocity ($v_{\rm crit}$), at which point the rotational and gravitational forces at the equator of the star are balanced.  A value often used to see how close to this point a star is the critical fraction $W=v/v_{\rm crit}$.

The variability and shape of the emission line has also been a matter of study: It can be a single or double peak \citep{Hanuschik1988} with variable values of asymmetry between the peaks \citep{Hanuschik1995vr} or have the shape of shell lines \citep{Hanuschik1995shell}. These profiles have been successfully modelled and theoretically reproduced \citep{Silaj2010}.

As well as showing emission line variability, the polarization of Be stars is also variable with time \citep{Coyne1967}.  The polarization is caused by radiation scattering due to free electrons \citep{Coyne1969} and is a function of the inclination angle.  As such the combination of spectroscopic and polarimetric observations can help constrain the true rotational velocity $v$. 

In this paper we present the first in a series of new analyses of the properties of a sample of Be stars \citep{Steele1999} that have been observed spectroscopically for more than two decades. The overall aim of the project is to understand the variability timescales of the objects and how that relates to the properties (mass, luminosity, rotational velocity) of the underlying B star.  For this first paper we concentrate on combining new polarimetric observations with the historical spectral data in order to understand the rotational velocity distribution of the sample.  This will then be used in future papers to explore the impact on line emission variability at various wavelengths.

In section \ref{sec:sampleselection} we describe the properties of the sample and previous work that has been carried out on it.  In section \ref{sec:data} we describe the acquisition and data reduction of our spectroscopic and polarimetric dataset and in section \ref{sec:halpha} make a basic classification of the appearance and variability of the H$\alpha$ line profile for each object.  Section \ref{sec:rotation} describes the calculation of projected rotational velocities and critical velocities for the sample.  Similarly, section \ref{sec:polarization} describes the correction  for each object of polarization degrees and angles for the effects of interstellar polarization.  In Section \ref{sec:analysis} we combine the results of the previous two sections to derive rotational velocities corrected for the effect of inclination angle.  Finally in section \ref{sec:conclusion} we make some concluding remarks.



\section{Sample selection} \label{sec:sampleselection}

Our sample is the same as the one introduced in \citet{Steele1999}. The 58 stars in the sample were were initially classified as Be stars in the database presented in \citet{JaschekEgret1982}, and were selected to cover several objects for each spectral and luminosity class for Be stars, to make it as representative as possible. However, because Be stars are variable, some of the stars classified as Be stars in 1982 might not show H$\alpha$ emission anymore. As the raw data used in their initial classification is generally unavailable and was compiled from a wide variety of sources, they may also have been misclassified. Several of the stars might also have changed to no emission at some point before or between the observations.

Previous work on the sample is described in a series of five papers published between 1999 and 2013. In paper I \citep{Steele1999}, the criteria for the selection of the sample was presented. A basic study of the rotational velocities of the stars was done based on a single spectrum of each object, as well as a reclassification of their spectral type. Paper II \citep{paper2} studied the same sample plus another 8 sources in the K band (2.05-2.22 $\mu$m). It determined that some objects do not show Br$\gamma$ emission ($\lambda=21660$ \AA), which might indicate no hydrogen emission in general, putting in doubt their classification as a Be star. Different sets of the spectra showed H I, He I, Mg II, Fe II and Na I lines, with He I and Mg II features as a good diagnostic of early spectral type. At the same time, stars that did not show Br$\gamma$ emission seemed to have a lower projected rotational velocity. Paper III \citep{paper3} studied 57 of the stars from paper 1 in the H band (1.53-1.69 $\mu$m). H I Brackett lines were again examined, as well as Fe II. It showed that the analysis of the H band spectra alone only allows for the classification into "early" (B0e-B4e) and "late" (B5e-B9e) types. Paper IV \citep{Howells2001} studied 52 of the stars from paper 1 with $JHK$ infrared photometry, separating values for the interstellar reddening and the circumstellar excess, and finding a strong correlation between the derived interstellar reddening values and the equivalent width of the interstellar sodium lines, giving confidence in the measured reddening. Paper V \citep{paper5} studied the equivalent width variability of the H$\alpha$ line for the measurements taken for 55 stars between 1998 and 2010, finding that stars of earlier types, with higher values of the projected rotational velocity, show a higher degree of variability in the H$\alpha$ emission.

\section{Data} \label{sec:data}

\subsection{Observations and Data Reduction}

The objects in our sample have been observed several times between 1998 and 2022 with multiple instruments. This work includes observations taken in 1998 and 2002 using the IDS spectrograph on the Isaac Newton Telescope (INT), 2009 observations with the medium resolution settings ($R\sim2500$) of the FRODOspec spectrograph on the Liverpool Telescope (LT) \citep{FRODOspec2004}, 2010 to 2021 observations with the high resolution ($R\sim5000$) FRODOspec settings, and 2022 observations with the low resolution ($R\sim350$) SPRAT spectrograph also on the LT \citep{SPRAT2014}. The years of observation, as well as the total number of observations, are shown in Table \ref{tab:observation_dates_list}.  All INT and LT data were re-reduced using the {\sc aspired} pipeline \citep{ASPIRED} for consistency.

In addition to these spectra, we have polarimetric observations taken with the MOPTOP polarimeter on the Liverpool Telescope \citep{MOPTOP2016, MOPTOP2018} in October 2022. Our sample is very bright, with apparent magnitudes in the range of $V=4.04$ and $V=10.60$ \citep{Steele1999}. For this reason, if we try to observe the brightest objects directly, the observation will be instantly saturated. To avoid this, we defocussed the telescope, with different values depending on how bright the stars are, allowing us to spread the light without saturating the detector. We also used the MOPTOP fast-rotation mode, which completes a cycle of 16 positions in 8 seconds, to avoid saturation as much as possible. The values used for the defocussing are shown in Table \ref{tab:defocus_table} to aid future observers of bright objects with the instrument.

Reduction of the MOPTOP data was carried out using aperture photometry routines from {\sc astropy} \citep{astropy}.  The extracted photometric counts were converted to Stokes $q$ and $u$ values following the procedure outlined in \cite{Shrestha2020}, with corrections for instrumental polarization made using the calibration values presented in \cite{moon}.

\begin{table} 
\centering
\caption{The values given for the defocus of the telescope when using MOPTOP for the different sources, depending on their apparent magnitude V.}
\label{tab:defocus_table}
    \begin{tabular}{cc}
        \hline
        V & Defocus (mm) \\
        \hline
        $m > 9.75$ & $0.0$ \\
        $9.75 > m > 6.75$ & $1.0$ \\
        $6.75 > m > 4.9$ & $2.0$ \\
        $4.9 > m > 4.04$ & $2.5$ \\
        \hline
    \end{tabular}
\end{table}

\subsection{Additional data}
As additional data, we have included the distances and the reddening. The distances have been calculated from parallax values obtained from the Gaia database \citep{GaiaDR3}, except in 3 cases where they have been taken from the Hipparcos catalog \citep{Hipparcos2007} because they were not available or the error of the Gaia values was larger than the actual values. The reddening $E(B-V)$ was obtained in \citet{Howells2001}, where we could find values for 52 of the stars in our data. These values have been added to Table \ref{tab:big_table_of_things}.


\section{H$\alpha$ classification} \label{sec:halpha}
\begin{figure}
	\includegraphics[width=\columnwidth]{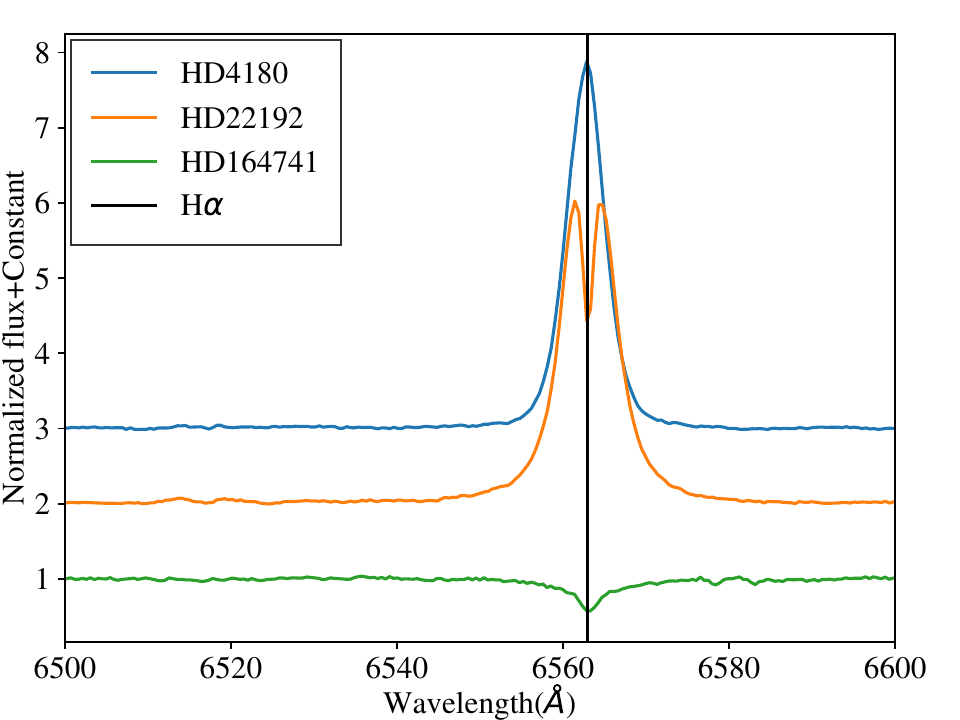}
   \caption{Representative plots of the three types of H$\alpha$ peaks in our sample, extracted from the 1998 observations. HD4180 is an example of a single peak, HD22192 is an example of a double peak, and HD164741 is instead an example of absorption in H$\alpha$.  The rest wavelength of H$\alpha$ is indicated by the vertical line.}
   \label{fig:spectra_peak_type}
\end{figure}

The most basic classification that we can make for Be stars from their spectra is taking into account the shape of the H$\alpha$ line: some Be stars will show a single peak emission, while others will show a mostly symmetric double peak emission. We show an example of this in Figure \ref{fig:spectra_peak_type}. At the same time, others will not show any H$\alpha$ emission anymore, and will show absorption of the line instead. Even more interesting, with a long time study like this one, we can also see some changing with time.  We will discuss this in a quantitative fashion in future papers.  However for now we simply classify them based on their overall appearance. In the most complicated cases of the ones with changing emission type, we will choose the most recent observation, as it is the most closely related to their polarization status.

\begin{figure}
	\includegraphics[width=\columnwidth]{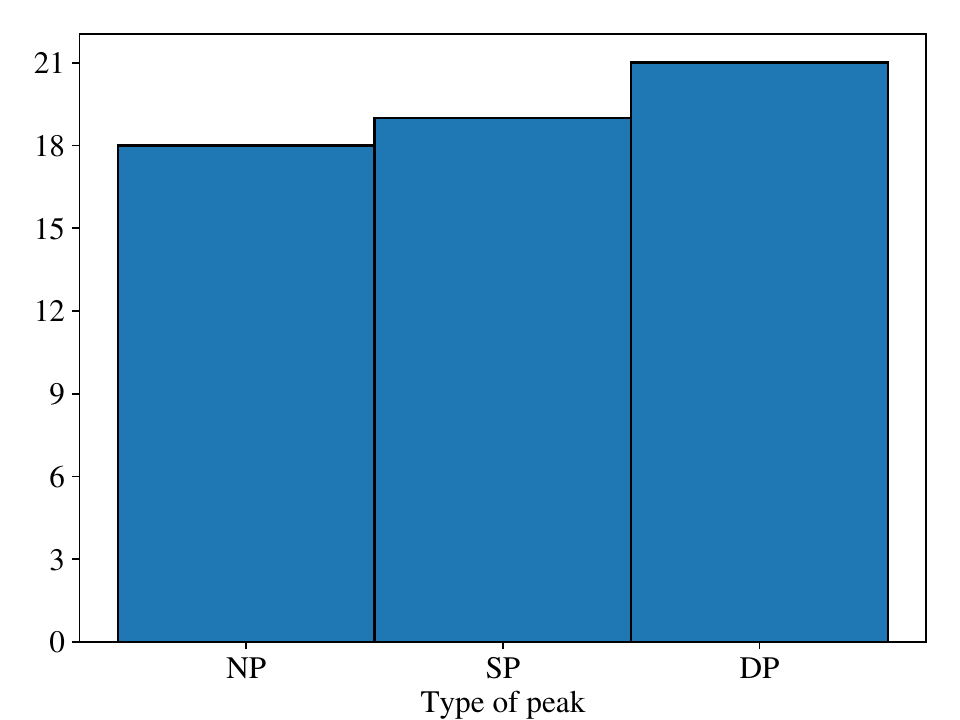}
   \caption{The distribution of our sample for the three different types of observed H$\alpha$, in this case, "NP" for "No peak", "SP" for "Single peak", and "DP" for "Double peak", classified from their last observation.}
   \label{fig:histogram_peak_type}
\end{figure}

From our sample of 58 stars, 10 never showed any H$\alpha$ emission between 1998 and our last observation. 16 show a single peak in all of our observations, while 21 show a double peak in all of them. The remaining 11 in our sample have changed with time, with 3 of them showing emission and 8 not showing emission in our last observation. In total, from our last observation, 18 do not show hydrogen emission and 40 show hydrogen emission lines (Figure \ref{fig:histogram_peak_type}). 

Examining the 11 objects that changed with time in more detail, we see that 9 of them at least spent some time as no emission, with another 2 always emitting and changing only between double peak and single peak. We can therefore assume a change rate between emission and absorption of 9/48 in 24 years, or 0.78\% of the total sample changing between emission and absorption every year (counting only the objects that we have observed actively changing from our first observation). Similarly if we instead assume all the objects that did not show any absorption in 1998 are changed from their original observations, compiled in \citet{JaschekEgret1982} but originally from the 1960s and 1970s, we would have a fraction 19/58 over an estimated time of 45 years (0.73\%/year). 

Previous determinations of this change rate \citep{McSwain2008,McSwain2009,paper5,Dimitrov2018} over shorter time periods (4--12 years) have ranged between 0.3 and 25 \%/year.  This wide range can be attributed to how the different samples used for the studies were constructed.
Using the same sample as here but only between 1998 and 2010 \citet{paper5} found 2 changing objects over 55 analyzed Be stars over a period of 12 years, or 0.30\%/year.
However \citep{McSwain2008} observed 16 Be stars in a sample of 191 B stars, with 11 of them changing between emission and no emission during a period of observation of 4 years. If these 11 are counted only from the 16 Be stars, then the change rate would be of 11/16 over 4 years, or 17\%/year.  Alternatively if we count them over the whole sample of B stars, then the change rate would be 11/191 over 4 years, or 1.4\%/year. A similar study \citet {McSwain2009} observed 45 Be stars in a sample of 296 B stars over 2 years, with 23 of them showing variability in emission. Counting only 23/45 for 2 years would give us a change rate of 25.5\%/year, while 23/296 for 2 years would give us instead a change rate of 3.89\%/year. Finally \citet{Dimitrov2018} observed 60 Be stars over a period on 3 years, and 4 of them changed in emission giving a change rate of 2.2\%/year. Overall comparing the values with at least similar methodologies (including our latest determination) it appears that a change rate of no more than a few percent per year is something like the true value.
 
\section{Results: rotation} \label{sec:rotation}

\begin{figure}
	\includegraphics[width=\columnwidth]{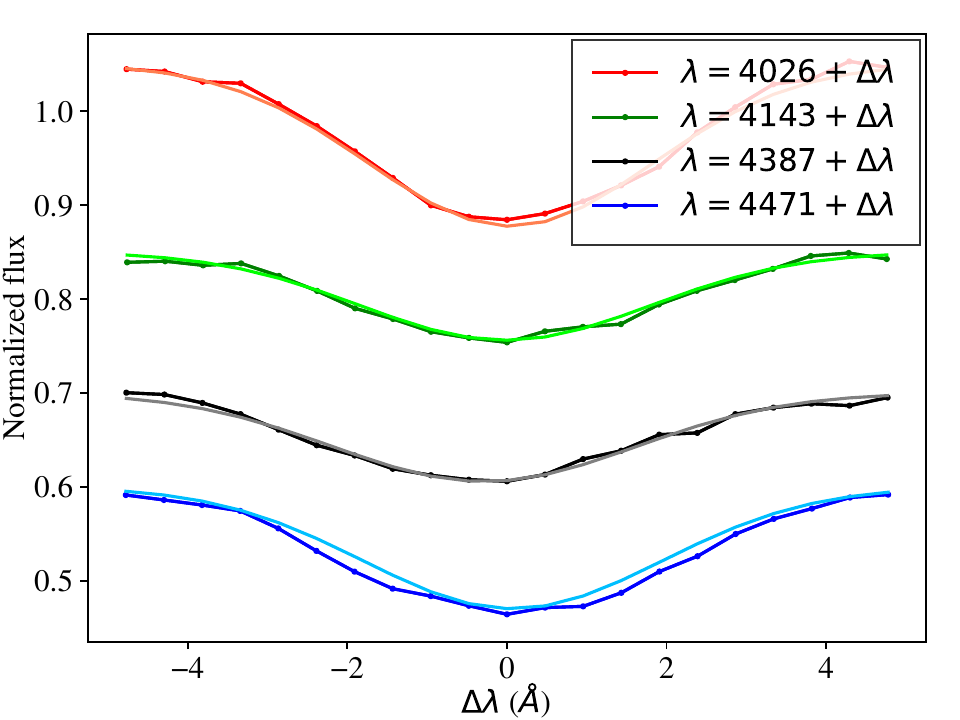}
   \caption{A plot, showing the fit to the helium I lines. In this case, it is the observation of star HD4180 taken in August 1998. We fit each line to a Gaussian profile and obtained the FWHM from them, and hence the velocity as described in Section \protect{\ref{sec:rotation}}. The results for each line in this example are vsini$(\lambda=4026$\AA$)=193.7 \pm 3.2$ km/s, vsini $(\lambda=4143$\AA$)=190.8 \pm 4.9$ km/s, vsini $(\lambda=4387$\AA$)=187.7 \pm 4.9$ km/s, vsini $(\lambda=4471$\AA$)=181.9 \pm 3.9$ km/s.}
   \label{fig:heI_lines_fit}
\end{figure}

\subsection{Calculation of the rotational velocities}
One of the most common values for Be stars is the rotational velocity of the central star, $vsini$. The way that we have chosen to obtain these values is using the He-{\sc I} absorption lines in our spectra, more precisely the lines at wavelengths 4026 \AA, 4143 \AA, 4387 \AA\ and 4471 \AA. We first fit a Gaussian profile to each of these lines, obtaining this way $\sigma$, the standard deviation of the Gaussian distribution, that then can be converted into the full width half maximum (FWHM) using:
\begin{equation}
    FWHM = 2 \sqrt{2 \ln 2} \sigma
\end{equation} 

\begin{figure}
	\includegraphics[width=\columnwidth]{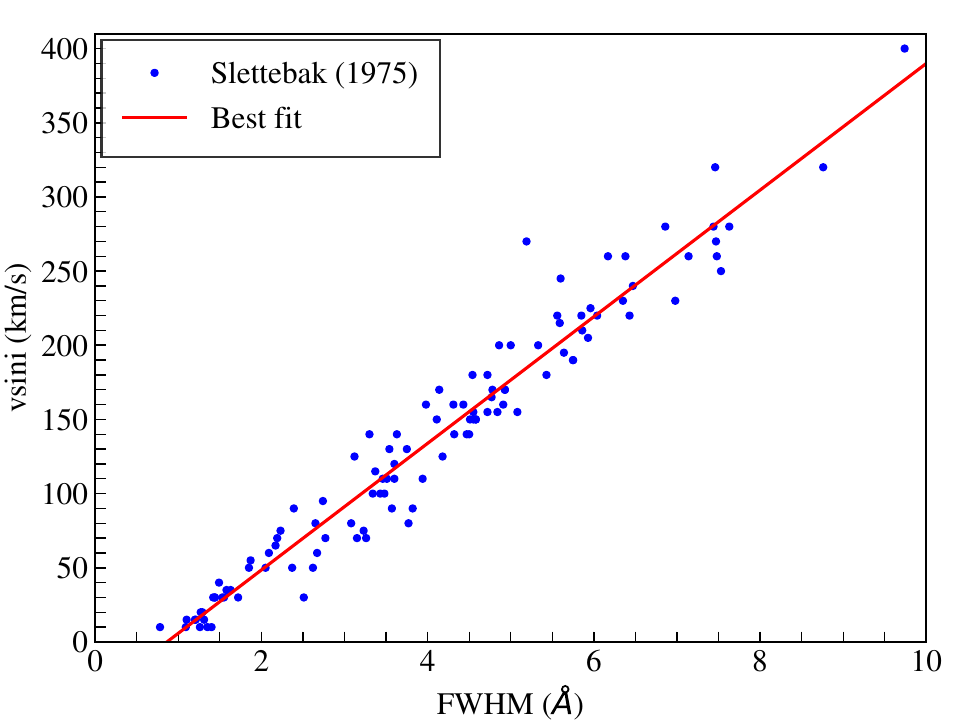}
   \caption{A plot of the rotational velocities of B stars vs the FWHM of the He I 4471 \AA line, extracted from tables 1 and 2 in \citet{Slettebak1975}. The results of the linear fit $y=Ax+B$ are $A=42.68\pm0.93$, $B=-36.84\pm4.17$, $R=0.9746$}
   \label{fig:slettebak_1975_fit_plot}
\end{figure}

These values of FWHM are not yet a measurement of the rotational velocity itself, since rotation is not the only line broadening mechanism.  We therefore convert them using the FWHM-$v \sin i$ relation derived from observational fits to rotationally broadened line profiles of model atmospheres of a wide sample of early type stars \citep{Slettebak1975}.
We will first do a linear fit for the values of FWHM and $v sin i$ of the He-{\sc I} line ($\lambda=4471$\AA) shown in the tables 1 and 2 of \citet{Slettebak1975}, as shown in Figure \ref{fig:slettebak_1975_fit_plot}, and then use this relation to convert our values to rotational velocities, applying a Doppler shift to the FWHM dependent slope to change the values to the other three near He-{\sc I} lines ($\lambda=4387$ \AA,$\lambda=4143$ \AA,$\lambda=4026$ \AA). The final equations obtained were:
\begin{equation}
\begin{split}
    v \sin i=42.68 F(\lambda=4471)-36.84 {\rm km s^{-1}} \\
    v \sin i=43.50 F(\lambda=4387)-36.84 {\rm km s^{-1}} \\
    v \sin i=46.06 F(\lambda=4143)-36.84 {\rm km s^{-1}} \\
    v \sin i=47.40 F(\lambda=4026)-36.84 {\rm km s^{-1}}
\end{split}
\end{equation}
with $F(\lambda)$ being the FWHM for the line at the corresponding wavelength.


\begin{figure}
	\includegraphics[width=\columnwidth]{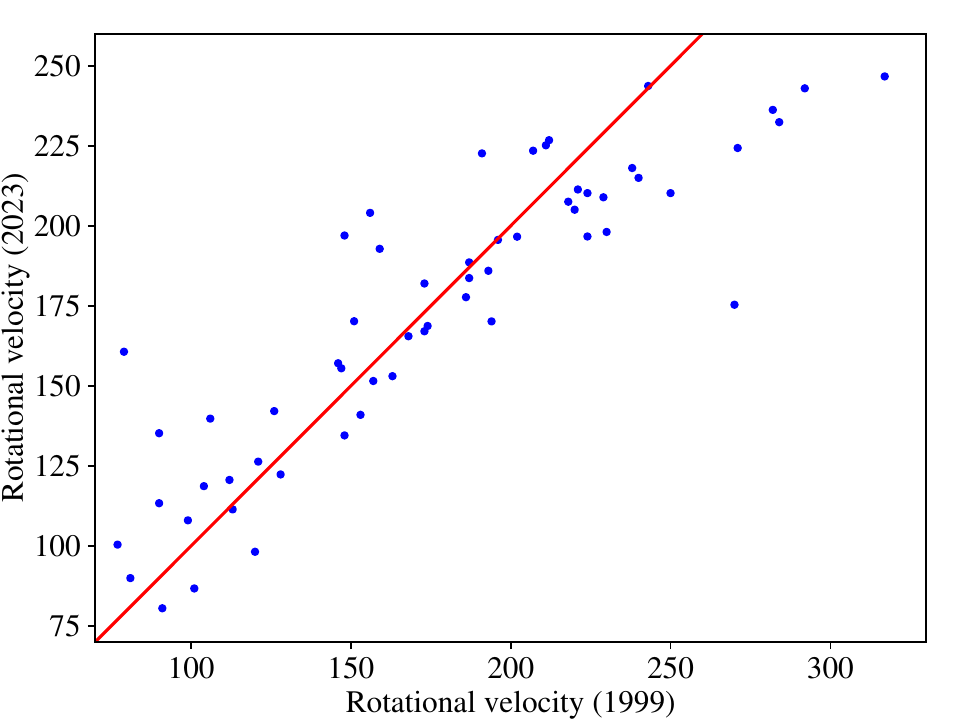}
   \caption{A plot comparing the values obtained for the rotational velocity in 2023 vs the values obtained for the rotational velocity in 1999. The red line is a 1:1 line for comparison that allows us to see that our rotational velocities seem to be similar when compared with the lower values of 1999, but lower when compared with the higher values of 1999. The cause of this difference might be that our results are the average of many years of observations, versus only one observation as in 1999.}
   \label{fig:2023_rot_vel_vs_1999_rot_vel_plot}
\end{figure}

For any particular object the results of the fits to these lines will show certain variation: a number of factors can cause this effect such as the noise in the spectrum or blending with another line. For that reason, we have chosen to calculated the weighted mean and standard deviation of the four lines and reject any values that are more than 4 $\sigma$ away from the average value.
With this, we have obtained a series of values in time for the rotational velocity of the star.  We do not expect any changes in the rotational velocity of a star over the span of just a few decades, so we take the weighted mean of these values for each object. These results are shown in Table \ref{tab:rot_vels_table} and represented in Figure \ref{fig:2023_rot_vel_vs_1999_rot_vel_plot}.

\begin{table} \footnotesize 
	\centering
	
 	\caption{\label{tab:rot_vels_table} List of observed objects, comparing their rotational velocities vs the rotational velocities obtained in \citet{Steele1999}}
	\begin{tabular}{llcc}
		\hline
        \multicolumn{1}{c}{Object} & \multicolumn{1}{c}{Other} & $v\sin i$ (km/s) & $v \sin i$  (km/s) \\
        & \multicolumn{1}{c}{Identifier} & (2023) & (1999) \\
        \hline
        CD-28 14778 & HD 171757 & 141 $\pm$ 23  & 153 $\pm$ 21 \\
        CD-27 11872 & HD 161103 & 210 $\pm$ 17  & 224 $\pm$ 33 \\
        CD-27 16010 & HD 214748 & 188.6 $\pm$ 7.2  & 187 $\pm$ 32 \\
        CD-25 12642 & HD 164741 & 100 $\pm$ 24  & 77 $\pm$ 18 \\
        CD-22 13183 & HD 172158 & 169 $\pm$ 79  & 174 $\pm$ 10 \\
        BD-20 5381 & HD 177015 & 197 $\pm$ 72  & 202 $\pm$ 10 \\
        BD-19 5036 & HD 170682 & 126 $\pm$ 17  & 121 $\pm$ 10 \\
        BD-12 5132 & HD 172252 & 98 $\pm$ 11  & 120 $\pm$ 43 \\
        BD-02 5328 & HD 196712 & 170.2 $\pm$ 8.4  & 151 $\pm$ 15 \\
        BD-01 3834 & HD 187350 & 166 $\pm$ 12  & 168 $\pm$ 34 \\
        BD-00 3543 & HD 173371 & 224 $\pm$ 15  & 271 $\pm$ 54 \\
        BD+02 3815 & HD 179343 & 197 $\pm$ 10  & 224 $\pm$ 14 \\
        BD+05 3704 & HD 168797 & 211 $\pm$ 59  & 221 $\pm$ 10 \\
        BD+17 4087 & HD 350559 & 204 $\pm$ 79  & 156 $\pm$ 39 \\
        BD+19 578 & HD 23016 & 215 $\pm$ 17  & 240 $\pm$ 70 \\
        BD+20 4449 & HD 191531 & 90 $\pm$ 43  & 81 $\pm$ 11 \\
        BD+21 4695 & HD 210129 & 157 $\pm$ 17  & 146 $\pm$ 10 \\
        BD+23 1148 & HD 250289 & 87 $\pm$ 77  & 101 $\pm$ 10 \\
        BD+25 4083 & HD 339483 & 161 $\pm$ 83  & 79 $\pm$ 11 \\
        BD+27 797 & HD 244894 & 197 $\pm$ 18  & 148 $\pm$ 74 \\
        BD+27 850 & HD 246878 & 121 $\pm$ 18  & 112 $\pm$ 25 \\
        BD+27 3411 & HD 183914 & 170 $\pm$ 13  & 194 $\pm$ 10 \\
        BD+28 3598 & HD 333452 & 113 $\pm$ 50  & 90 $\pm$ 12 \\
        BD+29 3842 & HD 333226 & 81 $\pm$ 20  & 91 $\pm$ 16 \\
        BD+29 4453 & HD 205618 & 247 $\pm$ 26  & 317 $\pm$ 20 \\
        BD+30 3227 & HD 171406 & 207.5 $\pm$ 8.3  & 218 $\pm$ 21 \\
        BD+31 4018 & HD 193009 & 225 $\pm$ 30  & 211 $\pm$ 11 \\
        BD+36 3946 & HD 228438 & 178 $\pm$ 23  & 186 $\pm$ 21 \\
        BD+37 675 & HD 18552 & 223 $\pm$ 21  & 207 $\pm$ 29 \\
        BD+37 3856 & HD 228650 & 119 $\pm$ 18  & 104 $\pm$ 17 \\
        BD+40 1213 & HD 33604 & 122.3 $\pm$ 7.4  & 128 $\pm$ 20 \\
        BD+42 1376 & HD 37657 & 195.6 $\pm$ 9.6  & 196 $\pm$ 10 \\
        BD+42 4538 & HD 216581 & 236 $\pm$ 30  & 282 $\pm$ 10 \\
        BD+43 1048 & HD 276738 & 205 $\pm$ 46  & 220 $\pm$ 20 \\
        BD+45 933 & HD 27846 & 135 $\pm$ 14  & 148 $\pm$ 16 \\
        BD+45 3879 & HD 211835 & 186 $\pm$ 10  & 193 $\pm$ 10 \\
        BD+46 275 & HD 6811 & 111 $\pm$ 20  & 113 $\pm$ 21 \\
        BD+47 183 & HD 4180 & 167 $\pm$ 12  & 173 $\pm$ 12 \\
        BD+47 857 & HD 22192 & 227 $\pm$ 12  & 212 $\pm$ 16 \\
        BD+47 939 & HD 25940 & 153.0 $\pm$ 8.0  & 163 $\pm$ 12 \\
        BD+47 3985 & HD 217050 & 232 $\pm$ 14  & 284 $\pm$ 20 \\
        BD+49 614 & HD 13867 & 140 $\pm$ 110  & 90 $\pm$ 27 \\
        BD+50 825 & HD 23552 & 184 $\pm$ 11 & 187 $\pm$ 10 \\
        BD+50 3430 & HD 207232 & 198 $\pm$ 19  & 230 $\pm$ 15 \\
        BD+51 3091 & HD 20551 & 139.8 $\pm$ 8.5  & 106 $\pm$ 10 \\
        BD+53 2599 & HD 203356 & 223 $\pm$ 22  & 191 $\pm$ 23 \\
        BD+55 552 & HD 13669 & 243 $\pm$ 29  & 292 $\pm$ 17 \\
        BD+55 605 & HD 14605 & 142 $\pm$ 25 & 126 $\pm$ 35 \\
        BD+55 2411 & HD 195554 & 192.8 $\pm$ 9.2  & 159 $\pm$ 90 \\
        BD+56 473 & V356 Per & 218 $\pm$ 23  & 238 $\pm$ 19 \\
        BD+56 478 & HD 13890 & 152 $\pm$ 30  & 157 $\pm$ 12 \\
        BD+56 484 & V502 Per & 182 $\pm$ 39  & 173 $\pm$ 16 \\
        BD+56 493 & - & 175 $\pm$ 41  & 270 $\pm$ 10 \\
        BD+56 511 & - & 108 $\pm$ 54  & 99 $\pm$ 14 \\
        BD+56 573 & - & 210 $\pm$ 51  & 250 $\pm$ 58 \\
        BD+57 681 & HD 237056 & 155 $\pm$ 20  & 147 $\pm$ 49 \\
        BD+58 554 & HD 237060 & 209 $\pm$ 17  & 229 $\pm$ 10 \\
        BD+58 2320 & HD 239758 & 244 $\pm$ 39  & 243 $\pm$ 20 \\
        \hline
	\end{tabular}
\end{table}

\subsection{Calculation of the critical velocity}

The critical velocity of a Be star is the theoretical equilibrium velocity at which the central B star would need to rotate over to eject the material on its equator purely by compensating gravity with rotation. We can find the formula for example in \citet{Townsend2004}:

\begin{equation} \label{eq:eq_3}
    v_{crit} = \sqrt{\frac{2}{3} \frac{G M} {R_P}}    
\end{equation}
with M the mass of the central star and $R_P$ its polar radius.

To obtain these values for our sample, we use classification of the spectral types from  \citet{Steele1999} (reproduced in our Table \ref{tab:big_table_of_things}) to take stellar luminosities from \citet{deJager1987, AstroQuantities2000}.  Masses and radii were then derived from those luminosity values using the equations shown in Tables 4 and 5 of \citet{Eker2018}.

First, we have calculated the masses from the luminosities, following the set of equations:

\begin{equation} \label{eq:eq_4}
\begin{split}
    &\log M = (\log L - 0.010)/4.329 ; 1.05<M/M_\odot\leq 2.40 \\
    &\log M = (\log L - 0.093)/3.967 ; 2.4<M/M_\odot\leq 7 \\
    &\log M = (\log L - 1.105)/2.865 ; 7<M/M_\odot \\
\end{split}
\end{equation}
calculating the three values and choosing the appropriate ones for their range.

From those values, and following the indications in \citet{Eker2018}, we have calculated as well the effective temperature:
\begin{equation} \label{eq:eq_5}
    \log T_{\mathrm{eff}} = -0.170 (\log M)^2 + 0.888 \log M + 3.671
\end{equation}

and then we finally calculate the radius of the star using Stefan-Boltzmann law:
\begin{equation} \label{eq:eq_6}
    R = \sqrt{\frac{L}{4\pi \sigma T_{\mathrm{eff}}^4}}
\end{equation}

After obtaining the values of the mass and radius of the star, we can calculate the $v_{crit}$ and the value of the critical fraction $W$, to show how close the obtained equatorial rotational velocity is to the critical value:
\begin{equation} \label{eq:crit_frac}
    W=\frac{v}{v_{crit}}
\end{equation}

We have also calculated the value of $Wsini$, where in equation \ref{eq:crit_frac} we used $vsini$ instead of $v$.
All of these values are listed in Table \ref{tab:big_table_of_things}.

\section{Results: polarization} \label{sec:polarization}
\subsection{Calculation of the polarization degree and angle}

After calculating the values for the rotational velocity ($v \sin i$), we move to the calculation of the polarization degree and angle of each band. We first calculate the Stokes parameters following the instructions for the double camera settings as shown in \citet{Shrestha2020}. After obtaining the Stokes parameters, we calculated the polarization degree and angle for each band according to the standard formulation. We have derived the polarization degree ($P$) and angle ($\theta$) values for the bands $V$, $B$ and $R$. These results can be found in Tables \ref{tab:pol_big_table_oct_2022} and \ref{tab:pol_big_table_oct_2022_pol_angles}.  As most of the values that we will compare our polarization values (section \ref{sec:analysis}) with are full-range/white-light instead of individual bands, we have calculated the average of the values q and u for the three bands and calculated the average values for the polarization degree ($\bar{P}$) and the angle ($\bar{\theta}$) from them.

We note here that MOPTOP polarimetry suffers from a systematic errors of $\sim 0.002$ in $P$ and $1^\circ$ in $\theta$ \citep{Shrestha2020}.  These systematic errors are not included in the values presented in Tables \ref{tab:pol_big_table_oct_2022} and \ref{tab:pol_big_table_oct_2022_pol_angles} but are incorporated into the final interstellar corrected values calculated in the next section.

\subsection{Separation of the interstellar and intrinsic polarization}
\begin{figure}
	\includegraphics[width=\columnwidth]{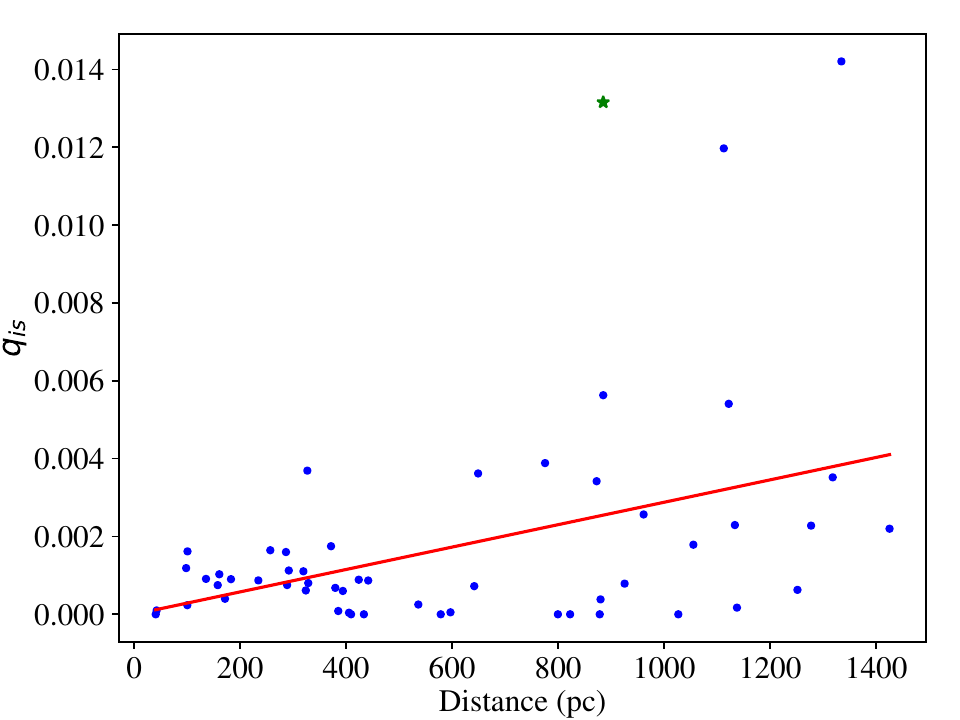}
   \caption{Here we see the polarization coefficient $q$ of nearby stars (in blue) compared to our object, BD+31 4018 (in green). The red line is a fit to the nearby star data.}
   \label{fig:interstellar_polarization_plot}
\end{figure}

We have attempted to separate interstellar polarization from the total polarization. To obtain the values for the interstellar polarization we have first selected stars from the 9286 object catalogue \citep{Heiles2000} within an angular distance of 5 degrees on the sky from each of our sample stars. We then found the distances to these group of stars using the Gaia catalog. We have separated the values of the Stokes parameters $q$ and $u$ from the catalogue polarization degree and angle for these stars, then performed a linear fit to them to find the values of interstellar $q$ and $u$ with respect the distance. Then we calculated $q_{is}$ and $u_{is}$ for our sample stars depending on their distance, using the previous linear fit, as shown in Figure \ref{fig:interstellar_polarization_plot}. The errors for the interstellar polarization are derived from the values of the linear fit.
We then subtracted the values of the interstellar $q$ and $u$ for our sample stars from the total values of their Stokes parameters, obtaining the intrinsic values ($q_{in}$ and $u_{in}$), and then calculated the values for both the interstellar and intrinsic polarization degree and angle as shown in Table \ref{tab:is_in_pol}. 

\section{Analysis} \label{sec:analysis}

\subsection{Comparison of $\bar{P}$ and $\bar{\theta}$ with other authors}

41 of the stars in our sample have had polarization measurements previously, so we can compare our values for $\bar{P}$ and $\bar{\theta}$ to the ones from previous authors. The sources for these reference values are \citet{Hall1958}, \citet{Behr1959}, \citet{Coyne1967}, \citet{Coyne1969}, \citet{Serkowski1968}, \citet{Serkowski1970}, \citet{McLean1976}, \citet{Coyne1976}, \citet{Poeckert1976}, \citet{McLean1978} and \citet{Yudin2001}. We have reduced these to 1950s, 1960s, 1970s and 2000s for simplicity, taking the average of all values where there is more than one.

\begin{table*}
    \caption{The values $\bar{P}$ and $\bar{\theta}$ shown here are the weighted average of their years. For the 1950s, we have used the ones shown in \citet{Hall1958} and \citet{Behr1959}, for the 1960s we have used the ones shown in \citet{Coyne1967}, \citet{Coyne1969} and \citet{Serkowski1968}, for the 1970 we have used \citet{Serkowski1970}, \citet{McLean1976}, \citet{Coyne1976}, \citet{Poeckert1976} and \citet{McLean1978}, for 2001 we have used the ones shown in \citet{Yudin2001}, and for 2022 our own calculated values. In bold, the values for the polarization angles that have shown a difference bigger than 10\degree \ over the past 70 years.}
	\label{tab:pol_values_history_comparison}
	\begin{tabular}{llccccccccccc}
		\hline
        \multicolumn{1}{c}{\multirow{2}{*}{Object}} & \multicolumn{1}{c}{\multirow{2}{*}{Other identifier}} & \multicolumn{2}{c}{1950s} & \multicolumn{2}{c}{1960s} & \multicolumn{2}{c}{1970s}& \multicolumn{2}{c}{2001}& \multicolumn{2}{c}{2022} & \multirow{2}{*}{Peak}\\ 

        & & $\bar{P}$ (\%)& $\bar{\theta}$ & $\bar{P}$ (\%)& $\bar{\theta}$ & $\bar{P}$ (\%)& $\bar{\theta}$ & $\bar{P}(\%)$ & $\bar{\theta}$ & $\bar{P}(\%)$ & $\bar{\theta}$\\
        \hline
       CD-27 11872 & HD 161103 &11.1&171&-&-&-&-&4.83&172&7.1&170 & SP \\
       CD-27 16010 & HD 214748 &-&-&-&-&1.75&163&0.09&155&0.4&169 & DP \\
       CD-25 12642 & HD 164741 &1.7&14&-&-&-&-&-&-&3.2&14 & NP \\
       CD-22 13183 & HD 172158 &1.7&178&-&-&-&-&4.65& 143&1.2&175 & SP \\
       BD-20 5381 & HD 177015 &1.1&0&-&-&-&-&0.51&0&1.3&179& DP \\
       BD-01 3834 & HD 187350 &1.7&82&-&-&-&-&0.78&82&0.4&80 & SP \\
       BD+05 3704 & HD 168797 &1.7&64&-&-&-&-&0.78&64&1.4&\bf{30} & Change  \\
       BD+19 578 & HD 23016 &-&-&-&-&1.48&87&0.60&87&0.5&87 & Change \\
       BD+29 4453 & HD 205618 &2.7&26&-&-&-&-&1.24&26&1.0&\bf{39} & SP \\
       BD+30 3227 & HD 171406 &0.6&13&-&-&-&-&0.28&13&1.6&\bf{65} & Change \\
       BD+31 4018 & HD 193009 &1.4&75&-&-&-&-&0.65&75&1.4&80 & SP \\
       BD+40 1213 & HD 33604 &2.8&168&-&-&-&-&1.29&168&1.5&165 & SP \\
       BD+42 1376 & HD 37657 &3.6&174&-&-&-&-&1.66&174&1.5&175 & Change\\
       BD+46 275 & HD 6811 & 2.08&90&-&-&8.75&94&7.4&90&0.9&89 & Change\\
       BD+47 183 & HD 4180 & 1.975 & 84 &1.05&84&0.815&82&0.70&85&1.1&82 & SP\\
       BD+47 857 & HD 22192 & 1.55&42&-&-&1.07&42&0.80&45&0.2&\bf{64} & DP \\
       BD+47 939 & HD 25940 &2.28&173&1.09&158&1.025&172&0.25&145&1.1&174 & SP \\
       BD+47 3985 & HD 217050 &3.9&75&1.65&69&1.123&71&1.55&75&0.4&\bf{53} & Change \\
       BD+55 605 & HD 14605 & 8.2&118&-&-&-&-&3.78&118&4.2&121 & Change\\
       BD+56 478 & HD 13890 & 7.5 & 107&-&-&-&-&3.46&107&3.7&112 & DP\\
        \hline
	\end{tabular}
\end{table*}

\begin{figure}
	\includegraphics[width=\columnwidth]{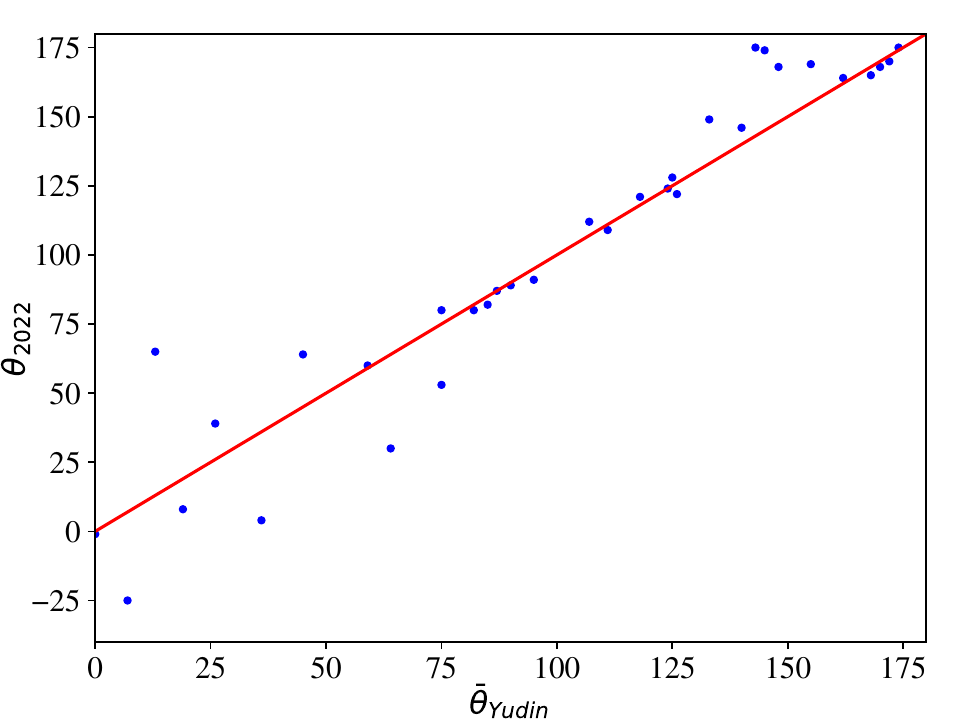}
   \caption{A plot of our (2022) polarization angles vs the polarization angles measured by Yudin (2001).}
   \label{fig:angles_plot_yudin_2022}
\end{figure}

In Table \ref{tab:pol_values_history_comparison} we compare our values with the most studied Be stars from our sample, for a historical perspective, while in Table \ref{tab:pol_values_history_comparison_yudin} we compare our entire sample with the ones from \citet{Yudin2001}. We can see that the polarization degree changes with time, and while the polarization angle is mostly constant (Figure \ref{fig:angles_plot_yudin_2022}), there are a number of objects with significant ($>10^\circ$) changes over the 70 year time baseline.  We highlight these objects in bold in Table \ref{tab:pol_values_history_comparison}.  We note that large changes in polarization angle are typically associated with large changes in the degree of polarization (either positive or negative).  However, we note that the objects that show such angle changes seem to be spread fairly evenly between those which do and do not show changes in the H$\alpha$ peak morphology (see last line of table).
Overall the changing polarization angle can be interpreted as either a varying ratio of interstellar and circumstellar polarization (which naturally have random polarization angles with respect to each other) as the disk strength varies, or as a change in the disk geometry as a function of time.  Such behaviour has, for example, been seen in the interferometric and polarimetric measurements presented by \citet{q97}.  We will explore this question on an object-by-object basis in a future paper.

\subsection{Removal of the inclination dependence}

As radiation escapes the Be star and is scattered through the disk, it will get polarized. This polarization has been modeled by \citet{Brown1977}:

\begin{equation}
    P \simeq \bar{\tau} (1 - 3 \gamma) \sin^2 i
\end{equation}
with P the polarization degree, $\bar{\tau}$ the average electron optical depth, $\gamma$ a shape factor that describes the asymmetry of the disk, and $i$ the inclination.

It is impossible for us to separate these values without knowing the value of the inclination. However, as we have measured previously the rotational velocity $v \sin i$, this allows us to obtain instead:
\begin{equation} \label{eq:pol_equation}
    \frac{P}{(v \sin i)^2} = \frac{K}{v^2}
\end{equation}
with $K=\bar{\tau}(1-3\gamma)$.

This allows us to derive a value for $v$ that is not dependent on the inclination of the star (although there is now an uncertainty related to the mean optical depth instead).  Similarly we can multiply the values shown in equation \ref{eq:pol_equation} by the critical velocity:

\begin{equation} \label{eq:pol_equation_2}
    \frac{P}{(v \sin i)^2} v_{\rm crit}^2 = \frac{K}{v^2} v_{\rm crit}^2 = \frac{K}{W^2}
\end{equation}

obtaining a measurement of the critical fraction $W$. 

\begin{figure*}
	\includegraphics[width=2\columnwidth]{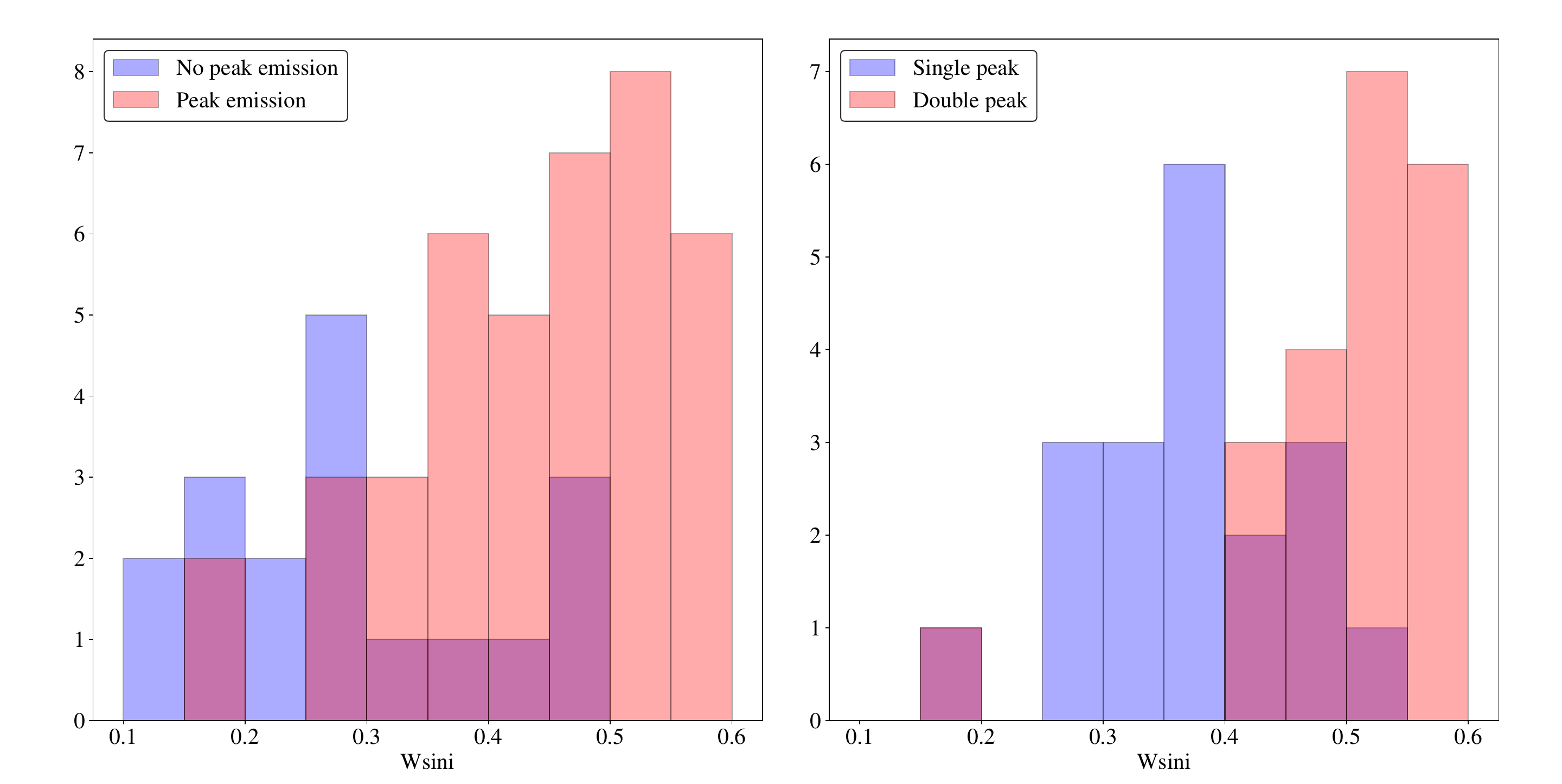}
   \caption{Two histograms showing the distribution of stars for different values and types of emission of the H$\alpha$ line with respect to $Wsini$.}
   \label{fig:histogram_plot_1}
\end{figure*}

\begin{figure*}
	\includegraphics[width=2\columnwidth]{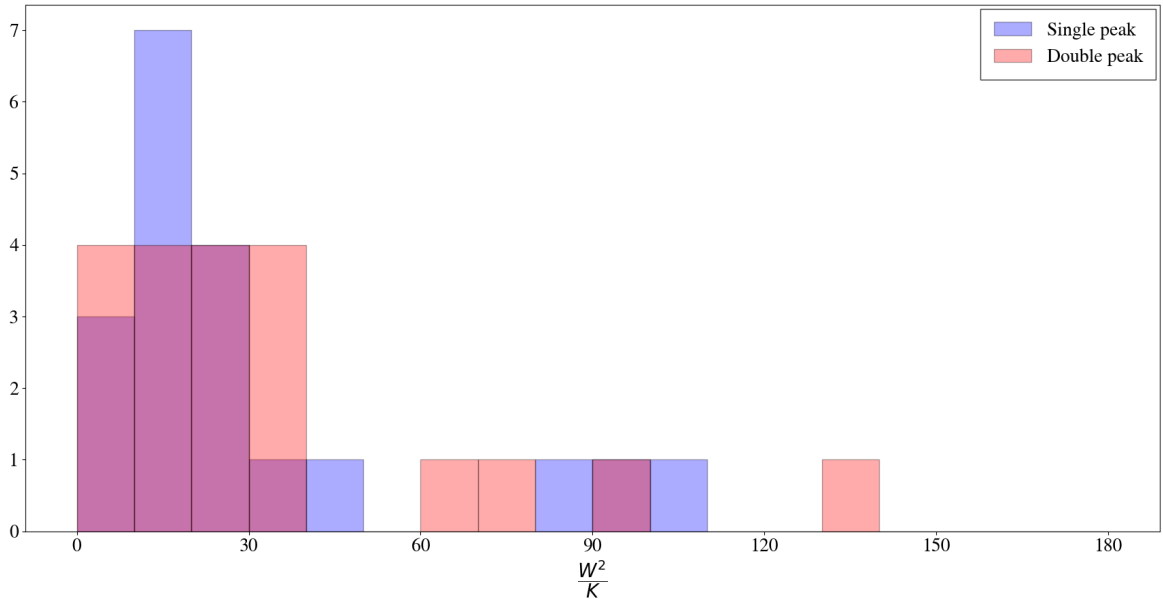}
   \caption{A histogram showing the distribution of stars for single and double peak H$\alpha$ line with respect to $\frac{W^2}{K}$.}
   \label{fig:histogram_plot_2}
\end{figure*}

\subsection{Testing the population differences}
In order to test the validity of the above procedure, we can compare the distribution of the various quantities $v \sin i$, $W \sin i$, $W^2/K$ and $v^2/K$ against the H$\alpha$ emission classification we made in Section \ref{sec:halpha}.  To do this we used Kolmogorov-Smirnov tests to compare the histograms of these quantities for the permutations of no- (NP), single peaked- (SP), and double peaked-emission DP).  The results of the tests are shown in Table \ref{tab:ks_test_results}.

\begin{table} 
	\centering
 	\caption{Results from the KS tests.  Distributions which are apparently drawn from the same parent population (i.e. $p>0.05$) are highlighted in bold. }
	\label{tab:ks_test_results}
	\begin{tabular}{ccccc}
		\hline
        Values & NP vs P & NP vs SP & NP vs DP & SP vs DP \\
        \hline
        $v \sin i$ & 0.00 & 0.04 & 0.00 & 0.00 \\
        $W \sin i$ & 0.00 & 0.01 & 0.00 & 0.00 \\
        $W^2/K$ & - & - & - &  0.64 \\
        $v^2/K$ & - & - & - &  0.98 \\
        \hline
	\end{tabular}
\end{table}

   
In all cases, when considering the distributions that include the $\sin i$ inclination term, the distributions for different H$\alpha$ emission properties all indicate a different parent population.

For example Figure \ref{fig:histogram_plot_1} (left) shows the distributions of P and NP objects are different in terms of $W \sin i$, with the first showing showing faster $W \sin i$ associated with $H\alpha$ emission, and the showing slower $W \sin i$ associated with $H\alpha$ absorption as expected from previous studies. 


Similarly the right panel shows the difference between single and double peaks, with double peaks seemingly rotating faster in terms of $W \sin i$. However, when we remove the angular dependence on the inclination ($W^2/K$) the null hypothesis can not be rejected \ref{fig:histogram_plot_2}. This can be interpreted as evidence for the idea that the cause of the difference between single and double peak H$\alpha$ emission is the inclination of the disk, as predicted by emission line modelling (e.g. \cite{Poeckert1976, poeckert1978}, \cite{sigut}). If that theory is taken as accepted, then our method of using polarization to eliminate the dependence on the inclination angle appears to work for the sample distribution.

\section{Conclusions} \label{sec:conclusion}

Using a sample of 58 stars previously classified as Be, we have used spectra taken over multiple epochs over a 24 year period to classify them based on their their qualitative $H\alpha$ line appearance. We classified these stars in three groups depending on whether they have emission in the shape of a single peak, in the shape of a double peak, or no emission.  We found a probability of $\sim 0.75$ percent per year that an object would change state between emission and non-emission.

We also calculated the projected rotational velocity ($v \sin i$) from multiple measurements of the He-{\sc I} line profiles. We then compared polarization measurements taken in 2022 with those from previous authors, revealing that the majority of objects show no significant changes in the polarization angle. This is to be expected if this were a mainly geometric parameter based on the system having a stable 3d space orientation.   However we do find a small sample of objects which seem to show genuine polarization angle variability which may be interpreted as either varying disk strength (optical depth) or geometry.  This will be investigated further in a future paper.

For each object we have used the degree of polarization to estimate a quantity that, when combined with the $v \sin i$, is related to the true equatorial rotational velocities.  We find a clear separation between Be stars with $H\alpha$ emission and stars that do not emit. Once the angular inclination factor is removed, the populations of single and double peak H$\alpha$ emitters are indistinguishable at least on a population basis.  In a future paper we will consider how to use quantitative measures of the variability of emission line properties of the objects in conjunction with this approach to constrain variability timescales in an inclination-independent fashion.


\section*{Acknowledgements}

We want to thank the Isaac Newton Group of Telescopes and the Liverpool Telescope for providing the measurements used for this publication. This paper makes use of data obtained from the Isaac Newton Group of Telescopes Archive which is maintained as part of the CASU Astronomical Data Centre at the Institute of Astronomy, Cambridge. The Liverpool Telescope is operated on the island of La Palma by Liverpool John Moores University in the Spanish Observatorio del Roque de los Muchachos of the Instituto de Astrofisica de Canarias with financial support from the UK Science and Technology Facilities Council under UKRI grant ST/T00147X/1. The authors want to thank Maurizio Salaris for his help with some of the calculations.
ACE thanks LJMU for financial support for his PhD studentship.

\section*{Data Availability}

All the data used for this publication is publicly available on the Isaac Newton Telescope and Liverpool Telescope data archives.




\bibliographystyle{mnras}
\bibliography{bibliography} 




\appendix

\section{Tables of data}
We have added here some tables that were too big to be inserted in the paper.

\begin{table*} \footnotesize 
\centering
\caption{The list of observations taken for the respective objects.}
\label{tab:observation_dates_list}
    \begin{tabular}{lccc}
        \hline
        \multicolumn{1}{c}{Object} & Alias & Observation dates & Total number of observations \\
        \hline
        CD-28 14778 & HD 171757 & 1998, 2010, 2018, 2019 & 6 \\
        CD-27 11872 & V3892 Sgr, HD 161103 & 1998, 2009, 2010, 2018, 2019, 2020 & 9\\
        CD-27 16010 & $\epsilon$ PsA, HR 8628, HD 214748 & 1998, 2009, 2010, 2013, 2018, 2019, 2020, 2022 & 15 \\
        CD-25 12642 & HD 164741 & 1998, 2009, 2010, 2018, 2019, 2020 & 8\\
        CD-22 13183 & HD 172158 & 1998, 2009, 2010 & 3\\
        BD-20 5381 & HD 177015 & 1998, 2009, 2010, 2019 & 5 \\
        BD-19 5036 & V3508 Sgr, HD 170682 & 1998, 2009, 2018, 2019 & 7\\
        BD-12 5132 & HD 172252 & 1998, 2018, 2019, 2020 & 5 \\
        BD-02 5328 & HD 196712 & 1998, 2009, 2010, 2013, 2018, 2019, 2022 & 12 \\
        BD-01 3834 & HD 187350 & 1998, 2009, 2010, 2018, 2019, 2020, 2022 & 11 \\
        BD-00 3543 & HD 173371 & 1998, 2009, 2010, 2018, 2019, 2020 & 9 \\
        BD+02 3815 & HD 179343 & 1998, 2009, 2010, 2018, 2019, 2020 & 10 \\
        BD+05 3704 & HD 168797 & 1998, 2002, 2009, 2010, 2018, 2019, 2020 & 10 \\
        BD+17 4087 & HD 350559 & 1998, 2018, 2019, 2020, 2022 & 9\\
        BD+19 578 & 13 Tau, HR 1126, HD 23016 & 1998, 2009, 2010, 2013, 2017, 2018, 2019, 2020, 2022 & 13\\
        BD+20 4449 & HD 191531 & 1998, 2009, 2010, 2022 & 5 \\
        BD+21 4695 & 25 Peg, HD 210129 & 1998, 2009, 2010, 2013, 2018, 2020, 2022 & 9 \\
        BD+23 1148 & HD 250289 & 1998, 2009, 2010, 2013, 2018, 2019, 2020, 2021, 2022 & 12 \\
        BD+25 4083 & HD 339483 & 1998, 2009, 2010, 2018, 2019, 2020, 2022 & 12\\
        BD+27 797 & HD 244894 & 1998, 2009, 2010, 2013, 2018, 2019, 2022 & 19 \\
        BD+27 850 & HD 246878 & 1998, 2009, 2010, 2013, 2018, 2020, 2021, 2022 & 12 \\
        BD+27 3411 & $\beta$2 Cyg, HR 7418, HD 183914 & 1998, 2002, 2009, 2010, 2018, 2019, 2020, 2022 & 12 \\
        BD+28 3598 & HD 333452 & 1998, 2009, 2010, 2018, 2019, 2020, 2022 & 7 \\
        BD+29 3842 & HD 333226 & 1998, 2009, 2010, 2018, 2019, 2020, 2022 & 9\\
        BD+29 4453 & HD 205618 & 1998, 2009, 2010, 2013, 2018, 2019, 2020, 2022 & 12 \\
        BD+30 3227 & HR 6971, HD 171406 & 1998, 2002, 2009, 2010, 2018, 2019, 2020, 2021, 2022 & 13 \\
        BD+31 4018 & V2113 Cyg, HD 193009 & 1998, 2009, 2010, 2013, 2018, 2019, 2020, 2022 & 13\\
        BD+36 3946 & HD 228438 & 1998, 2002, 2009, 2010, 2013, 2018, 2019, 2020, 2022 & 15 \\
        BD+37 675 & HR 894, HD 18552 & 1998, 2009, 2010, 2013, 2018, 2019, 2020, 2022 & 10 \\
        BD+37 3856 & HD 228650 & 1998, 2009, 2010, 2013, 2018, 2019, 2020, 2022 & 14 \\
        BD+40 1213 & HD 33604 & 1998, 2009, 2010, 2013, 2018, 2019, 2020, 2021, 2022 & 14 \\
        BD+42 1376 & V434 Aur, HD 37657 & 1998, 2009, 2010, 2013, 2018, 2019, 2020, 2021, 2022 & 13 \\
        BD+42 4538 & HD 216581 & 1998, 2002, 2009, 2010, 2013, 2018, 2019, 2020, 2022 & 12 \\
        BD+43 1048 & HD 276738 & 1998, 2009, 2010, 2013, 2018, 2019, 2022 & 11 \\
        BD+45 933 & HD 27846 & 1998, 2009, 2010, 2013, 2018, 2019, 2020, 2022 & 13 \\
        BD+45 3879 & HD 211835 & 1998, 2002, 2009, 2010, 2013, 2018, 2019, 2020, 2022 & 12 \\
        BD+46 275 & $\varphi$ And, HR 335, HD 6811 & 1998, 2009, 2010, 2013, 2018, 2019, 2020, 2022 & 15 \\
        BD+47 183 & 22 Cas, HR 193, HD 4180 & 1998, 2009, 2010, 2013, 2018, 2019, 2020, 2022 & 14 \\
        BD+47 857 & $\psi$ Per, HR 1087, HD 22192 & 1998, 2009, 2010, 2013, 2017, 2018, 2019, 2020, 2022 & 13 \\
        BD+47 939 & 48 Per, HR 1273, HD 25940 & 1998,  2001, 2009, 2010, 2013, 2017, 2018, 2019, 2020, 2022 & 15\\
        BD+47 3985 & EW Lac, HR 8731, HD 217050 & 1998, 2002, 2005, 2017, 2018, 2019, 2020, 2022 & 12 \\
        BD+49 614 & HD 13867 & 1998, 2009, 2010, 2013, 2018, 2022 & 9 \\
        BD+50 825 & HR 1160, HD 23552 & 1998, 2007, 2009, 2010, 2013, 2018, 2019, 2020, 2021, 2022 & 15\\
        BD+50 3430 & HD 207232 & 1998, 2009, 2010, 2013, 2018, 2019, 2020, 2022 & 15 \\
        BD+51 3091 & HR 8259, HD 20551 & 1998, 2009, 2010, 2013, 2018, 2019, 2022 & 12 \\
        BD+53 2599 & HD 203356 & 1998, 2009, 2010, 2013, 2018, 2019, 2020, 2022 & 14 \\
        BD+55 552 & HD 13669 & 1998, 2009, 2010, 2013, 2018, 2019, 2022 & 12 \\
        BD+55 605 & V361 Per, HD 14605 & 1998, 2009, 2010, 2013, 2018, 2022 & 8 \\
        BD+55 2411 & HD 195554 & 1998, 2002, 2009, 2010, 2013, 2018, 2019, 2020, 2022 & 17 \\
        BD+56 473 & V356 Per & 1998, 2009, 2010, 2013, 2018, 2022 & 7 \\
        BD+56 478 & V358 Per, HD 13890 & 1998, 2002, 2009, 2010, 2013, 2018, 2019, 2020, 2022 & 10 \\
        BD+56 484 & V502 Per & 1998, 2009, 2010, 2013, 2018, 2020, 2022 & 12 \\
        BD+56 493 & - & 1998, 2009, 2010, 2013, 2018, 2019, 2022 & 9\\
        BD+56 511 & - & 1998, 2009, 2010, 2013, 2018, 2019, 2021, 2022 & 11 \\
        BD+56 573 & - & 1998, 2009, 2010, 2013, 2018, 2019, 2022 & 12 \\
        BD+57 681 & HD 237056 & 1998, 2009, 2010, 2013, 2018, 2019, 2020, 2022 & 13 \\
        BD+58 554 & HD 237060 & 1998, 2009, 2010, 2013, 2018, 2019, 2020, 2022 & 12 \\
        BD+58 2320 & HD 239758 & 1998, 2009, 2010, 2013, 2018, 2019, 2020, 2022 & 12 \\
        \hline
    \end{tabular}
\end{table*}

\begin{table*} \footnotesize 
	\centering
 	\caption{The list of objects, including the spectral and luminosity classes, distances, theoretical luminosities, masses and radius of the star, theoretical critical velocity and reddening. $\dagger$ indicates that the distances are from the Hipparcos catalog instead of Gaia. All the theoretical $log (L/L_\odot)$ values come from \citet{deJager1987} and they only depend on the spectral type, while the $M/M_\odot$, $R/R_\odot$ and $v_{crit}$ are derived from equations \ref{eq:eq_3}, \ref{eq:eq_4}, \ref{eq:eq_5} and \ref{eq:eq_6}. The values of the reddening $E(B-V)$ come from \citet{Howells2001}.}
    \label{tab:big_table_of_things}
	\begin{tabular}{llccccccccc}
		\hline
        \multicolumn{1}{c}{Object}	&	\multicolumn{1}{c}{Spectral	Type}	&	d (pc) &	$log	(L/L_\odot)$	&	$M/M_\odot$	&	$R/R_\odot$	&	$v_{crit}$	(km/s) &	$E(B-V)$	\\
        \hline																				
        CD-28 14778	&	B2III	&	$1624	\pm	73$	&	4.047	&	10.638	&	5.490	&	496.1 &	0.54\\	
        CD-27 11872	&	B0.5V-III	&	$1270	\pm	38$	&	4.258	&	12.604	&	5.855	&	522.9	&	0.86\\	
        CD-27 16010	&	B8IV	&	$168.9	\pm	8.4$	&	2.337	&	3.678	&	2.844	&	405.3	&	0.16\\	
        CD-25 12642	&	B0.7III	&	$1414	\pm	39$	&	5.034	&	23.517	&	7.983	&	611.7	&	-\\	
        CD-22 13183	&	B7V	&	$882	\pm	38$	&	2.203	&	3.403 &	2.719	&	398.7	&	0.27\\	
        BD-20 5381	&	B5V	&	$643	\pm	16$	&	2.681	&	4.491	&	3.220	&	420.9	&	-\\	
        BD-19 5036	&	B4III	&	$667	\pm	16$	&	3.562	&	7.204	&	4.889	&	432.6	&	0.51\\	
        BD-12 5132	&	BN0.2III	&	$1630	\pm	190$	&	4.846	&	20.219	&	7.328	&	592.0	&	0.78\\	
        BD-02 5328	&	B7V	&	$350.4	\pm	6.6	$&	2.203	&	3.403	&	2.719	&	398.7	&	-0.06\\	
        BD-01 3834	&	B2IV	&	$2010	\pm	110$	&	3.84	&	9.008	&	5.197	&	469.2	&	0.08\\	
        BD-00 3543	&	B7V	&	$287.1	\pm	3.0$	&	2.203	&	3.403	&	2.719 &	398.7	&	0.06\\	
        BD+02 3815	&	B7-8sh	&	$790	\pm	510^\dagger$	&	2.094	&	3.195	&	2.624	&	393.2	&	-0.03\\	
        BD+05 3704	&	B2.5V	&	$418.8	\pm	7.7$	&	3.332	&	6.554	&	4.206	&	444.9	&	-0.13\\	
        BD+17 4087	&	B6III-V	&	$2096	\pm	64$	&	2.758	&	4.697	&	3.316	&	424.1	&	0.08\\	
        BD+19 578	&	B8V	&	$159.3	\pm	2.1$	&	1.985	&	2.999	&	2.536	&	387.6	&	-\\	
        BD+20 4449	&	B0III	&	$2130	\pm	180$	&	4.928	&	21.596	&	7.600	&	600.8	&	0.16\\	
        BD+21 4695	&	B6III-V	&	$204.5	\pm	3.7$	&	3.08	&	5.662	&	3.774	&	436.5	&	0.03\\	
        BD+23 1148	&	B2III	&	$2180	\pm	130$	&	4.047	&	10.638	&	5.490	&	496.1	&	0.89\\	
        BD+25 4083	&	B0.7-1III	&	$1269	\pm	22$	&	4.573	&	16.236	&	6.549	&	561.1	&	-\\	
        BD+27 797	&	B0.5V	&	$1977	\pm	58$	&	4.258	&	12.604	&	5.855	&	522.9	&	1.59\\	
        BD+27 850	&	B1.5IV	&	$1231	\pm	29$	&	4.081	&	10.933	&	5.544	&	500.4	&	0.65\\	
        BD+27 3411	&	B8V	&	$122.1	\pm	1.2$	&	1.985	&	2.999	&	2.536	&	387.6	&	-0.18\\	
        BD+28 3598	&	BO9II	&	$2295	\pm	71$	&	5.516	&	34.643	&	10.253	&	655.1	&	1.18\\	
        BD+29 3842	&	B1II	&	$4560	\pm	250$	&	4.93	&	21.631	&	7.607	&	601.0	&	0.73\\	
        BD+29 4453	&	B1.5V	&	$1224	\pm	48$	&	3.733	&	8.266	&	5.067	&	455.2	&	0.11\\	
        BD+30 3227	&	B4V	&	$364.5	\pm	9.1$	&	2.936	&	5.208	&	3.557	&	431.2	&	0.12\\	
        BD+31 4018	&	B1.5V	&	$885	\pm	18$	&	3.733	&	8.266	&	5.067	&	455.2	&	-0.36\\	
        BD+36 3946	&	B1V	&	$2350	\pm	140$	&	3.998	&	10.228	&	5.415	&	489.8	&	0.58\\	
        BD+37 675	&	B7V	&	$224.1	\pm	2.8$	&	2.203	&	3.403	&	2.719	&	398.7	&	0.13\\	
        BD+37 3856	&	B0.5V	&	$2195	\pm	56$	&	4.258	&	12.604	&	5.855	&	522.9	&	1.11\\	
        BD+40 1213	&	B2.5IV	&	$1102	\pm	33$	&	3.716	&	8.153	&	5.048	&	452.9	&	0.17\\	
        BD+42 1376	&	B2V	&	$698	\pm	16$	&	3.465	&	6.670	&	4.798	&	420.2	&	0.31\\	
        BD+42 4538	&	B2.5V	&	$594.6	\pm	9.9$	&	3.332	&	6.553	&	4.206	&	444.9	&	0.27\\	
        BD+43 1048	&	B6IIIsh	&	$971	\pm	27$	&	3.08	&	5.662	&	3.774	&	436.5	&	0.39\\	
        BD+45 933	&	B1.5V	&	$957	\pm	26$	&	3.733	&	8.266	&	5.067	&	455.2	&	0.51\\	
        BD+45 3879	&	B1.5V	&	$1876	\pm	73$	&	3.733	&	8.266	&	5.067	&	455.2	&	-\\	
        BD+46 275	&	B5III	&	$220	\pm	29^\dagger$	&	3.319	&	6.504	&	4.182	&	444.5	&	0.2\\	
        BD+47 183	&	B2.5V	&	$216	\pm	18^\dagger$	&	3.332	&	6.554	&	4.206	&	444.9	&	0.19\\	
        BD+47 857	&	B4V-III	&	$167.7	\pm	3.7$	&	2.936	&	5.208	&	3.557	&	431.2	&	0.37\\	
        BD+47 939	&	B2.5V	&	$156.8	\pm	7.2$	&	3.332	&  6.554	&	4.206	&	444.9	&	0.41\\	
        BD+47 3985	&	B1.5V	&	$287.4	\pm	5.8$	&	3.733	&	8.266	&	5.067	&	455.2	&	0.32\\	
        BD+49 614	&	B5III	&	$499	\pm	86$	&	3.319	&	6.504	& 4.182	&	444.5	&	0.23\\	
        BD+50 825	&	B7V	&	$249.8	\pm	2.1$	&	2.203	&	3.403	&	2.719	& 398.7	&	0.46\\	
        BD+50 3430	&	B8V	&	$384.8	\pm	4.4$	&	1.985	&	2.999	&	2.536	&	387.6	&	0.07\\	
        BD+51 3091	&	B7III	&	$457.7	\pm	6.3$	&	2.576	&	4.226	&	3.096	&	416.3	&	0.13\\	
        BD+53 2599	&	B8V	&	$637	\pm	23$	&	1.985	&	2.999	&	2.536	&	387.6 &	-\\	
        BD+55 552	&	B4V	&	$699	\pm	11$	&	2.936	&	5.208	&	3.557	&	431.2	&	0.07\\	
        BD+55 605	&	B1V	&	$2311	\pm	84$	&	3.998	&	10.227	&	5.415	&	489.8	&	0.78\\	
        BD+55 2411	&	B8.5V	&	$305.7	\pm	9.2$	&	1.884	&	2.828	&	2.459	&	382.2	&	0.27\\	
        BD+56 473	&	B1V-III	&	$2460	\pm	98$	&	3.998	&	10.228	&	5.415	&	489.8	&	0.57\\	
        BD+56 478	&	B1.5V	&	$2100	\pm	230$	&	3.733	&	8.266	&	5.067	&	455.2	&	0.6\\	
        BD+56 484	&	B1V	&	$2381	\pm	90$	&	3.998	&	10.228	&	5.415	&	489.8	&	0.78\\	
        BD+56 493	&	B1V	&	$2850	\pm	160$	&	3.998	&	10.228	&	5.415	&	489.8	&	0.57\\	
        BD+56 511	&	B1III	&	$2590	\pm	120$	&	4.508	&	15.409	&	6.389	&	553.5	&	0.72\\	
        BD+56 573	&	B1.5V	&	$2200	\pm	120$	&	3.733	&	8.266	&	5.067	&	455.2	&	0.89\\	
        BD+57 681	&	B0.5V	&	$1132	\pm	33$	&	4.258	&	12.604	&	5.855 &	522.9	&	1.17\\	
        BD+58 554	&	B7V	&	$662.2	\pm	6.9$	&	2.203	&	3.403	&	2.719	&	398.7	&	0.27\\	
        BD+58 2320	&	B2V	&	$1254	\pm	24$	&	3.465	&	6.670	&	4.798	&	420.2	&	0.62\\
    \hline
	\end{tabular}
\end{table*}

\begin{table*} 
	\centering
 	\caption{The list of objects, with the calculated polarization degrees for each BVR band, and the weighted polarization average of them, for the Oct 2022 observations. The errors quoted in this table have been calculated only from photon counting statistics and do not include the MOPTOP systematic errors of $\sim0.002$. Because the error from the calculations is smaller than the systematic error, the final value of the error for the average polarization degree is $S(\bar{P})=0.002$. Final polarization measurements incorporating the systematic error and the effect of interstellar polarization are given in Table \ref{tab:is_in_pol}.}
    \label{tab:pol_big_table_oct_2022}
	\begin{tabular}{lccccccc}
		\hline
        \multicolumn{1}{c}{Object} & $P_V$ & S($P_V$) & $P_B$ & S($P_B$) & $P_R$ & S($P_R$) & $\bar{P}$  \\
        \hline
        CD-28 14778  &  0.03467  &  $6\cdot 10^{-5}$  &  0.0367  &  0.0001  &  0.02589  &  $3\cdot 10^{-5}$  & 0.032  \\
        CD-27 11872  &  0.07152  &  $3\cdot 10^{-5}$  &  0.07755  &  $6\cdot 10^{-5}$  &  0.06398  &  $2\cdot 10^{-5}$  & 0.071  \\
        CD-27 16010  &  0.0035203  &  $8\cdot 10^{-7}$  &  0.005  &  0.003  &  0.0025063  &  $3\cdot 10^{-7}$  & 0.004  \\
        CD-25 12642  &  0.03264  &  $5\cdot 10^{-5}$  &  0.0420  &  0.0001  &  0.02167  &  $3\cdot 10^{-5}$  & 0.032  \\
        CD-22 13183  &  0.010014  &  $9\cdot 10^{-6}$  &  0.0142  &  0.0002  &  0.01086  &  $2\cdot 10^{-5}$  & 0.012  \\
        BD-20 5381  &  0.0162  &  0.0001  &  0.0122  &  0.0004  &  0.00977  &  $3\cdot 10^{-5}$  & 0.013  \\
        BD-19 5036  &  0.006470  &  $2\cdot 10^{-6}$  &  0.0442  &  $4\cdot 10^{-5}$  &  0.03798  &  $1\cdot 10^{-5}$  & 0.029 \\
        BD-12 5132  &  0.106  &  0.001  &  0.106  &  0.001  &  0.064  &  0.001  & 0.091 \\
        BD-02 5328  &  0.003889  &  $3\cdot 10^{-6}$  &  0.0046  &  0.0003  &  0.0036205  &  $9\cdot 10^{-07}$  & 0.004 \\
        BD-01 3834  &  0.003  &  0.001  &  0.0040  &  0.0003  &  0.00516  &  $6\cdot 10^{-5}$  & 0.004 \\
        BD-00 3543  &  0.008073  &  $6\cdot 10^{-6}$  &  0.008  &  0.005  &  0.006470  &  $2\cdot 10^{-6}$  & 0.007 \\
        BD+02 3815  &  0.010120  &  $8\cdot 10^{-6}$  &  0.01210  &  $2\cdot 10^{-5}$  &  0.007586  &  $3\cdot 10^{-6}$  & 0.010  \\
        BD+05 3704  &  0.005612  &  $1\cdot 10^{-6}$  &  0.009922  &  $8\cdot 10^{-6}$  &  0.030378  &  $6\cdot 10^{-6}$  & 0.014  \\
        BD+17 4087  &  0.0221  &  0.0003  &  0.02  &  0.02  &  0.0206  &  0.0002  & 0.022 \\
        BD+19 578  &  0.004838  &  $1\cdot 10^{-6}$  &  0.0055  &  0.0001  &  0.0046032  &  $7\cdot 10^{-7}$  & 0.005  \\
        BD+20 4449  &  0.01052  &  $2\cdot 10^{-5}$  &  0.01025  &  $8\cdot 10^{-5}$  &  0.007306  &  $9\cdot 10^{-6}$  & 0.009  \\
        BD+21 4695  &  0.002800  &  $3\cdot 10^{-6}$  &  0.004  &  0.004  &  0.001353  &  $1\cdot 10^{-6}$  & 0.002 \\
        BD+23 1148  &  0.03  &  0.08  &  0.0316  &  0.0003  &  0.0311  &  0.0009  & 0.032 \\
        BD+25 4083  &  0.016  &  0.001  &  0.02  &  0.01  &  0.02  &  0.06  & 0.016 \\
        BD+27 797  &  0.02  &  0.03  &  -  &  -  &  0.02  &  0.01  & 0.017  \\
        BD+27 850  &  0.0099  &  0.0001  &  0.010  &  0.001  &  0.00595  &  $5\cdot 10^{-5}$  & 0.009  \\
        BD+27 3411  &  0.0040326  &  $7\cdot 10^{-7}$  &  0.005223  &  $2\cdot 10^{-6}$  &  0.020104  &  $1\cdot 10^{-6}$  & 0.010  \\
        BD+28 3598  &  0.0538  &  0.0002  &  0.0633  &  0.0003  &  0.037  &  0.009  & 0.051  \\
        BD+29 3842  &  0.0167  &  0.0001  &  0.0202  &  0.0004  &  0.01389  &  $5\cdot 10^{-5}$  & 0.017 \\
        BD+29 4453  &  0.01030  &  $3\cdot 10^{-5}$  &  0.0104  &  0.0002  &  0.00929  &  $1\cdot 10^{-5}$  & 0.010 \\
        BD+30 3227  &  0.01325  &  $1\cdot 10^{-5}$  &  0.0247  &  0.0004  &  0.009893  &  $7\cdot 10^{-6}$  & 0.016  \\
        BD+31 4018  &  0.0140  &  0.0009  &  0.01481  &  $4\cdot 10^{-5}$  &  0.01329  &  $2\cdot 10^{-5}$  & 0.014  \\
        BD+36 3946  &  0.00725  &  $5\cdot 10^{-5}$  &  0.0062  &  0.0002  &  0.00746  &  $1\cdot 10^{-5}$  & 0.007  \\
        BD+37 675  &  0.00427  &  $2\cdot 10^{-5}$  &  0.00553  &  $4\cdot 10^{-5}$  &  0.004496  &  $6\cdot 10^{-6}$  & 0.005  \\
        BD+37 3856  &  0.0154  &  0.0001  &  0.0124  &  0.0005  &  0.01304  &  $7\cdot 10^{-5}$  & 0.014 \\
        BD+40 1213  &  0.016  &  0.006  &  0.015  &  0.002  &  0.0146  &  0.0002  & 0.015  \\
        BD+42 1376  &  0.0156  &  0.0002  &  0.02  &  0.01  &  0.02  &  0.02  & 0.015  \\
        BD+42 4538  &  0.00922  &  $3\cdot 10^{-5}$  &  0.0093  &  0.0003  &  0.00704  &  $2\cdot 10^{-5}$  & 0.008 \\
        BD+43 1048  &  0.027  &  0.004  &  0.024  &  0.003  &  0.024  &  0.002  & 0.025  \\
        BD+45 933  &  0.0229  &  0.0001  &  0.02446  &  $8\cdot 10^{-5}$  &  0.0230  &  0.0002  & 0.023 \\
        BD+45 3879  &  0.00539  &  $3\cdot 10^{-5}$  &  0.0064  &  0.0002  &  0.00665  &  $1\cdot 10^{-5}$  & 0.006 \\
        BD+46 275  &  0.009412  &  $1\cdot 10^{-6}$  &  0.00966  &  $3\cdot 10^{-5}$  &  0.0089211  &  $6\cdot 10^{-7}$  & 0.009 \\
        BD+47 183  &  0.011340  &  $1\cdot 10^{-6}$  &  0.010746  &  $3\cdot 10^{-6}$  &  0.0098539  &  $7\cdot 10^{-7}$  & 0.011 \\
        BD+47 857  &  0.0045151  &  $4\cdot 10^{-7}$  &  0.00568  &  $3\cdot 10^{-5}$  &  0.0032336  &  $1\cdot 10^{-7}$  & 0.002 \\
        BD+47 939  &  0.0115475  &  $3\cdot 10^{-7}$  &  0.0120  &  0.0002  &  0.0109207  &  $1\cdot 10^{-7}$  & 0.011 \\
        BD+47 3985  &  0.004066  &  $2\cdot 10^{-6}$  &  0.004237  &  $5\cdot 10^{-6}$  &  0.0024482  &  $8\cdot 10^{-7}$  & 0.004 \\
        BD+49 614  &  0.01308  &  $5\cdot 10^{-5}$  &  0.01  &  0.14  &  0.01148  &  $2\cdot 10^{-5}$  & 0.012 \\
        BD+50 825  &  0.006995  &  $2\cdot 10^{-6}$  &  0.00843  &  $4\cdot 10^{-5}$  &  0.006553  &  $1\cdot 10^{-6}$  & 0.007 \\
        BD+50 3430  &  0.00559  &  $1\cdot 10^{-5}$  &  0.01  &  0.01  &  0.004803  &  $4\cdot 10^{-6}$  & 0.005 \\
        BD+51 3091  &  0.007973  &  $6\cdot 10^{-6}$  &  0.00772  &  $2\cdot 10^{-5}$  &  0.006755  &  $3\cdot 10^{-6}$  & 0.007 \\
        BD+53 2599  &  0.00668  &  $4\cdot 10^{-5}$  &  0.007  &  0.002  &  0.00465  &  $2\cdot 10^{-5}$  & 0.006 \\
        BD+55 552  &  0.03  &  0.01  &  0.0297  &  0.0004  &  0.02  &  0.01  & 0.027  \\
        BD+55 605  &  0.04  &  0.01  &  0.0460  &  0.0006  &  0.04  &  0.02  & 0.042 \\
        BD+55 2411  &  0.001685  &  $6\cdot 10^{-6}$  &  0.00424  &  $4\cdot 10^{-5}$  &  0.001212  &  $2\cdot 10^{-6}$  & 0.002 \\
        BD+56 473  &  0.0485  &  0.0006  &  0.0501  &  0.0002  &  0.043  &  0.008  & 0.047 \\
        BD+56 478  &  0.0386  &  0.0007  &  0.0392  &  0.0002  &  0.03  &  0.03  & 0.037  \\
        BD+56 484  &  -  &  -  &  0.0417  &  0.0006  &  0.0327  &  0.0006  & 0.037  \\
        BD+56 493  &  0.046  &  0.008  &  0.0475  &  0.0005  &  0.040  &  0.002  & 0.044  \\
        BD+56 511  &  0.041  &  0.001  &  0.0437  &  0.0003  &  0.04  &  0.14  & 0.041  \\
        BD+56 573  &  0.04  &  0.03  &  0.0475  &  0.0006  &  0.036  &  0.003  & 0.042 \\
        BD+57 681  &  0.05971  &  $2\cdot 10^{-5}$  &  0.06348  &  $6\cdot 10^{-5}$  &  0.05678  &  $1\cdot 10^{-5}$  & 0.060  \\
        BD+58 554  &  0.0325  &  0.0002  &  0.0346  &  0.0002  &  0.032  &  0.001  & 0.033 \\
        BD+58 2320  &  0.0201  &  0.0007  &  0.02  &  0.04  &  0.0162  &  0.0002  & 0.017  \\
    \hline
	\end{tabular}
\end{table*}

\begin{table*} 
	\centering
 	\caption{The list of objects, with the calculated polarization angles for each BVR band, and the weighted polarization angle average of them, for the Oct 2022 observations. The angle values are between 0 and 180\degree. The errors quoted in this table have been calculated only from photon counting statistics and do not include the MOPTOP systematic errors of $\sim1^\circ$. Because the error from the calculations is smaller than the systematic error, the final value of the error for the average polarization degree is $S(\bar{\theta})=1^\circ$. Final polarization measurements incorporating the systematic error and the effect of interstellar polarization are given in Table \ref{tab:is_in_pol}.}
    \label{tab:pol_big_table_oct_2022_pol_angles}
	\begin{tabular}{lccccccc}
		\hline
        \multicolumn{1}{c}{Object} & $\theta_V$ & S($\theta_V$) & $\theta_B$ & S($\theta_B$) & $\theta_R$ & S($\theta_R$) & $\bar{\theta}$ \\
        \hline
        CD-28 14778 & 184.97 & 0.04 & 184.93 & 0.07 & 181.62 & 0.02 & 4  \\
        CD-27 11872 & 169.86 & 0.01 & 170.67 & 0.03 & 170.052 & 0.009 & 170 \\
        CD-27 16010 & 172.87 & 0.05 & 164 & 7 & 172.89 & 0.02 & 169 \\
        CD-25 12642 & 191.84 & 0.03 & 193.46 & 0.05 & 196.29 & 0.03 & 14  \\
        CD-22 13183 & 174.011 & 0.005 & 179.9 & 0.1 & 169.74 & 0.04 & 175  \\
        BD-20 5381 & 182.2 & 0.2 & 177.2 & 0.6 & 177.41 & 0.06 & 179  \\
        BD-19 5036 & 176.200 & 0.003 & 190.06 & 0.02 & 191.528 & 0.007 & 10  \\
        BD-12 5132 & 169.07 & 0.03 & 169.07 & 0.02 & 162.5 & 0.1 & 168  \\
        BD-02 5328 & 257.16 & 0.02 & 284 & 53 & 268.998 & 0.002 & 91  \\
        BD-01 3834 & 268 & 191 & 163 & 18 & 260.9 & 0.4 & 80  \\
        BD-00 3543 & 185.47 & 0.02 & 186 & 3 & 176.200 & 0.003 & 3 \\
        BD+02 3815 & 179.84 & 0.03 & 180.23 & 0.08 & 178.28 & 0.01 & 0 \\
        BD+05 3704 & 281.803 & 0.003 & 286.00 & 0.01 & 218.018 & 0.005 & 30  \\
        BD+17 4087 & 186 & 3 & 189 & 1231 & 188 & 5 & 8 \\
        BD+19 578 & 259.407 & 0.004 & 275 & 2 & 265.906 & 0.002 & 87 \\
        BD+20 4449 & 190.21 & 0.08 & 191.0 & 0.4 & 188.05 & 0.02 & 10  \\
        BD+21 4695 & 157.5 & 0.3 & 134 & 47 & 149.57 & 0.09 & 155  \\
        BD+23 1148 & 162 & 62 & 164.8 & 0.2 & 164.0 & 0.5 & 164 \\
        BD+25 4083 & 345 & 3 & 259 & 27 & 343 & 63 & 166  \\
        BD+27 797 & 187 & 25 & - & - & 189 & 17 & 8  \\
        BD+27 850 & 147.1 & 0.9 & 155 & 4 & 142.1 & 0.4 & 149 \\
        BD+27 3411 & 331.280 & 0.006 & 246.54 & 0.02 & 332.0686 & 0.0007 & 153 \\
        BD+28 3598 & 218.73 & 0.05 & 219.54 & 0.08 & 221 & 3 & 40 \\
        BD+29 3842 & 260.6 & 0.2 & 260.5 & 0.5 & 259.00 & 0.07 & 80  \\
        BD+29 4453 & 39.14 & 0.02 & 39.4 & 0.2 & 38.94 & 0.01 & 39  \\
        BD+30 3227 & 244.82 & 0.02 & 334.8 & 0.4 & 246.98 & 0.02 & 65  \\
        BD+31 4018 & 259.1 & 0.8 & 261.90 & 0.03 & 257.30 & 0.02 & 80  \\
        BD+36 3946 & 266 & 2 & 257 & 6 & 342.77 & 0.06 & 168  \\
        BD+37 675 & 144.2 & 0.3 & 149 & 2 & 148.33 & 0.05 & 147  \\
        BD+37 3856 & 307.5 & 0.2 & 306.0 & 0.8 & 307.6 & 0.1 & 127  \\
        BD+40 1213 & 162 & 5 & 166 & 2 & 166.9 & 0.4 & 165 \\
        BD+42 1376 & 172.6 & 0.4 & 176 & 19 & 176 & 51 & 175 \\
        BD+42 4538 & 81.7 & 0.2 & 92 & 1 & 83.0 & 0.1 & 86  \\
        BD+43 1048 & 356 & 1 & 90.7 & 0.4 & 178 & 2 & 88  \\
        BD+45 933 & 318.1 & 0.4 & 318.8 & 0.4 & 318 & 1 & 138  \\
        BD+45 3879 & 27.0 & 0.2 & 21.7 & 0.8 & 25.12 & 0.03 & 24  \\
        BD+46 275 & 85.205 & 0.002 & 91.30 & 0.02 & 90.0299 & 0.0009 & 89 \\
        BD+47 183 & 81.48 & 0.02 & 81.8 & 0.6 & 82.590 & 0.009 & 82 \\
        BD+47 857 & 38.778 & 0.001 & 54.9 & 0.3 & 47.7664 & 0.0004 & 64  \\
        BD+47 939 & 354.3414 & 0.0003 & 175.51 & 0.04 & 352.6755 & 0.0001 & 174  \\
        BD+47 3985 & 58.11 & 0.06 & 44.23 & 0.06 & 55.27 & 0.04 & 53 \\
        BD+49 614 & 102.87 & 0.03 & 106 & 25 & 103.123 & 0.007 & 104 \\
        BD+50 825 & 326.338 & 0.002 & 328.02 & 0.05 & 322.430 & 0.001 & 146  \\
        BD+50 3430 & 189.9 & 0.1 & 191 & 19 & 9.62 & 0.02 & 10  \\
        BD+51 3091 & 24.57 & 0.07 & 22.3 & 0.6 & 19.93 & 0.02 & 22  \\
        BD+53 2599 & 37.3 & 0.6 & 113 & 52 & 40.5 & 0.3 & 33 \\
        BD+55 552 & 110 & 9 & 114.3 & 0.4 & 110 & 3 & 112  \\
        BD+55 605 & 119 & 15 & 123.6 & 0.8 & 119 & 13 & 121  \\
        BD+55 2411 & 141.5 & 0.2 & 146.8 & 0.4 & 310.8 & 0.3 & 149  \\
        BD+56 473 & 108.0 & 0.1 & 110.50 & 0.05 & 109 & 1 & 109  \\
        BD+56 478 & 110.7 & 0.4 & 114.0 & 0.1 & 111 & 13 & 112  \\
        BD+56 484 & - & - & 126.8 & 0.9 & 123.6 & 0.7 & 124 \\
        BD+56 493 & 120 & 8 & 123.5 & 0.6 & 121 & 2 & 122  \\
        BD+56 511 & 111.8 & 0.4 & 115.8 & 0.1 & 113 & 36 & 114  \\
        BD+56 573 & 118 & 36 & 120.0 & 0.6 & 119 & 3 & 119  \\
        BD+57 681 & 307.438 & 0.002 & 309.860 & 0.005 & 307.1085 & 0.0005 & 128 \\
        BD+58 554 & 296.2 & 0.2 & 298.0 & 0.1 & 296.8 & 0.7 & 117 \\
        BD+58 2320 & 58 & 1 & 67 & 34 & 55.6 & 0.5 & 60  \\
    \hline
	\end{tabular}
\end{table*}

\begin{table*} 
	\centering
 	\caption{Comparison of our values with the ones from \citet{Yudin2001}.}
	\label{tab:pol_values_history_comparison_yudin}
	\begin{tabular}{llcccccc}
		\hline
        \multicolumn{1}{c}{Object} & \multicolumn{1}{c}{Object (2001)} & $vsini$ (2001) (km/s)& $P$ (\%) (2001)& $\theta$ (2001)& $vsini$ (2022) (km/s) & $\bar{P}$ (2022) & $\bar{\theta}$ (2022) \\
        \hline
    CD-28 14778 & HD 171757	&	$182\pm18$	&	$0.43\pm0.10$	&	36	&	141	$\pm$	23	&	3.2	&	4	\\
    CD-27 11872 & HD 161103	&	$260\pm26$	&	$4.83\pm0.12$	&	172	&	210	$\pm$	17	&	7.1	&	170	\\
    CD-27 16010 & HD 214748	&	$222\pm26$	&	$0.09\pm0.10$	&	155	&	189	$\pm$	7	&	0.4	&	169	\\
    CD-22 13183 & HD 172158	&	$232\pm26$	&	$0.54\pm0.10$	&	143	&	169	$\pm$	79	&	1.2	&	175	\\
    BD-20 5381 & HD 177015	&	$236\pm24$	&	$0.51\pm0.20$	&	0	&	197	$\pm$	72	&	1.3	&	179	\\
    BD-19 5036 & HD 170682	&	$121\pm12$	&	-	&	-	&	126	$\pm$	17	&	2.9	&	10	\\
    BD-12 5132 & HD 172252	&	$146\pm15$	&	$4.65\pm0.20$	&	148	&	98	$\pm$	11	&	9.1	&	168	\\
    BD-02 5328 & HD 196712	&	$199\pm19$	&	$0.35\pm0.04$	&	95	&	170	$\pm$	8	&	0.4	&	91	\\
    BD-01 3834 & HD 187350	&	$199\pm20$	&	$0.78\pm0.20$	&	82	&	166	$\pm$	12	&	0.4	&	80	\\
    BD-00 3543 & HD 173371	&	$291\pm~4$	&	-	&	-	&	224	$\pm$	15	&	0.7	&	3	\\
    BD+02 3815 & HD 179343	&	$261\pm36$	&	-	&	-	&	197	$\pm$	10	&	1.0	&	0	\\
    BD+05 3704 & HD 168797	&	$251\pm~6$	&	$0.78\pm0.20$	&	64	&	211	$\pm$	59	&	1.4 &	30	\\
    BD+17 4087 & HD 350559	&	$186\pm19$	&	-	&	-	&	204	$\pm$	79	&	2.2 &	8	\\
    BD+19 578 & HD 23016	&	$278\pm14$	&	$0.60\pm0.04$	&	87	&	215	$\pm$	17	&	0.5	&	87	\\
    BD+20 4449 & HD 191531	&	$~80\pm~3$	&	-	&	-	&	90	$\pm$	43	&	0.9	&	10  \\
    BD+21 4695 & HD 210129	&	$201\pm32$	&	$0.28\pm0.04$	&	7	&	157	$\pm$	17	&	0.2	&	155	\\
    BD+23 1148 & HD 250289	&	$125\pm13$	&	$2.95\pm0.18$	&	162	&	87	$\pm$	77	&	3.2	&	164	\\
    BD+27 797 & HD 244894	&	$177\pm18$	&	$1.61\pm0.18$	&	19	&	197	$\pm$	18	&	1.7	&	8	\\
    BD+27 850 & HD 246878	&	$137\pm14$	&	$0.55\pm0.18$	&	133	&	121	$\pm$	18	&	0.9 &	149	\\
    BD+27 3411 & HD 183914	&	$220\pm25$	&	-	&	-	&	170	$\pm$	13	&	1.0	&	153	\\
    BD+29 4453 & HD 205618	&	$299\pm65$	&	$1.24\pm0.20$	&	26	&	247	$\pm$	26	&	1.0	 &	39	\\
    BD+30 3227 & HD 171406	&	$264\pm10$	&	$0.28\pm0.20$	&	13	&	208	$\pm$	8	&	1.6	&	65 \\
    BD+31 4018 & HD 193009	&	$220\pm26$	&	$0.65\pm0.20$	&	75	&	225	$\pm$	30	&	1.4	&	80	\\
    BD+36 3946 & HD 228438	&	$219\pm22$	&	$1.15\pm0.20$	&	170	&	178	$\pm$	23	&	0.7	&	168	\\
    BD+37 675 & HD 18552	&	$260\pm12$	&	-	&	-	&	223	$\pm$	21	&	0.5	&	147	\\
    BD+40 1213 & HD 33604	&	$134\pm20$	&	$1.29\pm0.20$	&	168	&	122 $\pm$	7	&	1.5	&	165	\\
    BD+42 1376 & HD 37657	&	$216\pm20$	&	$1.66\pm0.20$	&	174	&	196	$\pm$	10	&	1.5 &	175	\\
    BD+42 4538 & HD 216581	&	$282\pm28$	&	-	&	-	&	236	$\pm$	30	&	0.8	&	86	\\
    BD+43 1048 & HD 276738	&	$220\pm22$	&	-	&	-	&	205	$\pm$	46	&	2.5	&	88	\\
    BD+45 933 & HD 27846	&	$134\pm15$	&	-	&	-	&	135	$\pm$	14	&	2.3	&	138	\\
    BD+45 3879 & HD 211835	&	$207\pm15$	&	-	&	-	&	186	$\pm$	10	&	0.6	&	24	\\
    BD+46 275 & HD 6811	    &	$~88\pm13$	&	$0.74\pm0.12$	&	90	&	111	$\pm$	20	&	0.9	&	89	\\
    BD+47 183 & HD 4180     &	$221\pm~9$	&	$0.70\pm0.13$	&	85	&	167	$\pm$	12	&	1.1	&	82	\\
    BD+47 857 & HD 22192	&	$297\pm41$	&	$0.80\pm0.15$	&	45	&	227	$\pm$	12	&	0.2	&	64	\\
    BD+47 939 & HD 25940	&	$198\pm31$	&	$0.25\pm0.10$	&	145	&	153	$\pm$	8	&	1.1	&	174	\\
    BD+47 3985 & HD 217050	&	$301\pm16$	&	$1.55\pm0.20$	&	75	&	232	$\pm$	14 &	0.4	&	53	\\
    BD+49 614 & HD 13867	&	$~93\pm16$	&	-	&	-	&	135	$\pm$	114	&	1.2	&	104	\\
    BD+50 825 & HD 23552	&	$218\pm~3$	&	$0.55\pm0.05$	&	140	&	184	$\pm$	11	&	0.7	&	146	\\
    BD+50 3430 & HD 207232	&	$281\pm28$	&	-	&	-	&	198	$\pm$	19	&	0.5	&	10	\\
    BD+55 552 & HD 13669	&	$304\pm22$	&	-	&	-	&	243	$\pm$	29	&	2.7	&	112	\\
    BD+55 605 & HD 14605	&	$150\pm~3$	&	$3.78\pm0.20$	&	118	&	142	$\pm$	25	&	4.2	&	121	\\
    BD+55 2411 & HD 195554	&	$215\pm21$	&	-	&	-	&	193	$\pm$	9	&	0.2	&	149	\\
    BD+56 473 & V356 Per	&	$258\pm26$	&	$4.38\pm0.18$	&	111	&	218	$\pm$	23	&	4.7	&	109	\\
    BD+56 478 & HD 13890	&	$179\pm~8$	&	$3.46\pm0.18$	&	107	&	152	$\pm$	30	&	3.7	&	112	\\
    BD+56 484 & V502 Per	&	$228\pm24$	&	$3.46\pm0.18$	&	124	&	182	$\pm$	39	&	3.7	&	124	\\
    BD+56 493 & BD+56 493	&	$311\pm31$	&	$3.64\pm0.18$	&	126	&	175	$\pm$	41	&	4.4	&	122	\\
    BD+56 511 & BD+56 511	&	$111\pm11$	&	-	&	-	&	108	$\pm$	54	&	4.1 &	114	\\
    BD+56 573 & BD+56 573   &	$335\pm34$	&	-	&	-	&	210	$\pm$	51	&	4.2 &	119	\\
    BD+57 681 & HD 237056	&	$160\pm17$	&	$5.90\pm0.18$	&	125	&	155	$\pm$	20	&	6.0	&	128	\\
    BD+58 2320 & HD 239758	&	$281\pm28$	&	$2.12\pm0.18$	&	59	&	244	$\pm$	39	&	1.7	&	60	\\

        \hline
	\end{tabular}
\end{table*}

\begin{table*} 
	\centering
 	\caption{The distance D to the object, and the minimum and maximum distance (d) that we take. The stars included in this range are used for our linear fits to obtain an estimation of the interstellar polarization. The large variation in these ranges is due to two main reasons: loss of linearity at big distances away from the star, or lack of data covering any other suitable range.}
	\label{tab:star_range}
	\begin{tabular}{lcccc}
		\hline
        \multicolumn{1}{c}{Object} & D (pc)& S(D) (pc) & Min d (pc) & Max d (pc) \\
        \hline
CD-28 14778	&	1625	&	73	&	78	&	4926	\\
CD-27 11872	&	1270	&	38	&	93	&	3657	\\
CD-27 16010	&	168.9	&	8.4	&	50.4	&	1268.6	\\
CD-25 12642	&	1414	&	39	&	698	&	2395	\\
CD-22 13183	&	882	&	38	&	43	&	997	\\
BD-20 5381	&	643	&	16	&	74	&	4714	\\
BD-19 5036	&	667	&	16	&	108	&	3898	\\
BD-12 5132	&	1630	&	190	&	43	&	597	\\
BD-02 5328	&	350.4	&	6.6	&	40.7	&	1425.2	\\
BD-01 3834	&	2010	&	110	&	49	&	1822	\\
BD-00 3543	&	287.1	&	3	&	42.5	&	4714.2	\\
BD+02 3815	&	790	&	510	&	132	&	2700	\\
BD+05 3704	&	418.8	&	7.7	&	46.3	&	779.9	\\
BD+17 4087	&	2096	&	64	&	131	&	3986	\\
BD+19 578	&	159.3	&	2.1	&	145.7	&	4405.5	\\
BD+20 4449	&	2130	&	180	&	131	&	381	\\
BD+21 4695	&	204.5	&	3.7	&	46.8	&	434.4	\\
BD+23 1148	&	2180	&	130	&	50	&	1263	\\
BD+25 4083	&	1269	&	22	&	93	&	4693	\\
BD+27 797	&	1977	&	58	&	854	&	3344	\\
BD+27 850	&	1231	&	29	&	54	&	374	\\
BD+27 3411	&	122.1	&	1.2	&	163.4	&	350.4	\\
BD+28 3598	&	2295	&	71	&	72	&	3527	\\
BD+29 3842	&	4560	&	250	&	126	&	1753	\\
BD+29 4453	&	1224	&	48	&	13	&	201	\\
BD+30 3227	&	364.5	&	9.1	&	43.4	&	597.3	\\
BD+31 4018	&	885	&	18	&	2005	&	3455	\\
BD+36 3946	&	2350	&	140	&	2005	&	3779	\\
BD+37 675	&	224.1	&	2.8	&	46.4	&	2005.2	\\
BD+37 3856	&	2195	&	56	&	126	&	4464	\\
BD+40 1213	&	1102	&	33	&	107	&	2449	\\
BD+42 1376	&	698	&	16	&	2007	&	4924	\\
BD+42 4538	&	594.6	&	9.9	&	37	&	963.7	\\
BD+43 1048	&	971	&	27	&	115	&	1879	\\
BD+45 933	&	957	&	26	&	2005	&	3346	\\
BD+45 3879	&	1876	&	73	&	51	&	1910	\\
BD+46 275	&	220	&	29	&	79	&	4446	\\
BD+47 183	&	216	&	18	&	102	&	162	\\
BD+47 857	&	167.7	&	3.7	&	147.6	&	4926.9	\\
BD+47 939	&	156.8	&	7.2	&	27.6	&	1224	\\
BD+47 3985	&	287.4	&	5.8	&	53.7	&	406.2	\\
BD+49 614	&	499	&	86	&	56	&	4578	\\
BD+50 825	&	249.8	&	2.1	&	2005.3	&	3455.3	\\
BD+50 3430	&	384.8	&	4.4	&	31.8	&	583.7	\\
BD+51 3091	&	457.7	&	6.3	&	24.5	&	582.2	\\
BD+53 2599	&	637	&	23	&	127	&	4254	\\
BD+55 552	&	699	&	11	&	43	&	3812	\\
BD+55 605	&	2311	&	84	&	101	&	4446	\\
BD+55 2411	&	305.7	&	9.2	&	111.2	&	515.9	\\
BD+56 473	&	2460	&	98	&	44	&	515	\\
BD+56 478	&	2100	&	230	&	38	&	964	\\
BD+56 484	&	2381	&	90	&	62	&	698	\\
BD+56 493	&	2850	&	160	&	28	&	96	\\
BD+56 511	&	2590	&	120	&	41	&	800	\\
BD+56 573	&	2200	&	120	&	2005	&	3963	\\
BD+57 681	&	1132	&	33	&	1527	&	4945	\\
BD+58 554	&	662.2	&	6.9	&	38.1	&	963.7	\\
BD+58 2320	&	1254	&	24	&	74	&	4714	\\
        \hline
	\end{tabular}
\end{table*}

\begin{table*} \footnotesize 
	\centering
 	\caption{The calculated values for the interstellar and intrinsic polarization degrees and angles. The estimated systematic error from MOPTOP is $S(\bar{P})=0.002$ and $S(\bar{\theta})=1$ degree. The error of the interstellar polarization values has been obtained from the linear fit to nearby stars, while the error of the intrinsic polarization values has been obtained from the calculations of $q_{in}$ and $u_{in}$ and the following derivation of $P_{in}$ and $\theta_{in}$.}
	\label{tab:is_in_pol}
	\begin{tabular}{lcccccccccc}
		\hline
\multicolumn{1}{c}{Object} & $\bar{P}$ & $\bar{\theta}$ & $P_{is}$ & $S(P_{is})$ & $\theta_{is}$ & $S(\theta_{is})$ & $P_{in}$ & $S(P_{in})$ & $\theta_{in}$ & $S(\theta_{in})$ \\
        \hline
CD-28 14778	&	0.032	&	4	&	0.0063	&	0.0005	&	23.72	&	0.05	&	0.01	&	0.01	&	171	&	3	\\
CD-27 11872	&	0.071	&	170	&	0.019	&	0.001	&	24.54	&	0.03	&	0.010	&	0.02	&	145.7	&	0.5	\\
CD-27 16010	&	0.004	&	169	&	0.0195	&	0.0005	&	34.92	&	0.01	&	0.015	&	0.005	&	16.7	&	0.3	\\
CD-25 12642	&	0.032	&	14	&	0.013	&	0.001	&	29.66	&	0.04	&	0.020	&	0.04	&	166	&	2	\\
CD-22 13183	&	0.012	&	175	&	0.0277	&	0.0009	&	23.17	&	0.02	&	0.001	&	0.007	&	141.6	&	0.9	\\
BD-20 5381	&	0.013	&	179	&	0.0079	&	0.0005	&	31.15	&	0.02	&	0.010	&	0.004	&	1	&	4	\\
BD-19 5036	&	0.029	&	10	&	0.00072	&	0.00005	&	42.312	&	0.004	&	0.009	&	0.005	&	25.9	&	0.3	\\
BD-12 5132	&	0.091	&	168	&	0.0044	&	0.0001	&	24.95	&	0.02	&	0.006	&	0.005	&	28.6	&	0.4	\\
BD-02 5328	&	0.004	&	91	&	0.0069	&	0.0004	&	34.24	&	0.01	&	0.011	&	0.005	&	176	&	2	\\
BD-01 3834	&	0.004	&	80	&	0.020	&	0.001	&	23.58	&	0.03	&	0.01	&	0.01	&	132.8	&	0.1	\\
BD-00 3543	&	0.007	&	3	&	0.029	&	0.003	&	28.85	&	0.04	&	0.02	&	0.02	&	135.45	&	0.01	\\
BD+02 3815	&	0.010	&	0	&	0.011	&	0.002	&	35.72	&	0.05	&	0.01	&	0.01	&	150.9	&	0.7	\\
BD+05 3704	&	0.014	&	30	&	0.0065	&	0.0003	&	27.88	&	0.02	&	0.007	&	0.005	&	31.9	&	0.2	\\
BD+17 4087	&	0.022	&	8	&	0.00094	&	0.00009	&	44.8314	&	0.0004	&	0.071	&	0.007	&	9.6	&	0.2	\\
BD+19 578	&	0.005	&	87	&	0.00083	&	0.00007	&	44.7854	&	0.0004	&	0.032	&	0.006	&	13.3	&	0.3	\\
BD+20 4449	&	0.009	&	10	&	0.002	&	0.002	&	20.6	&	0.7	&	0.014	&	0.006	&	25.6	&	0.2	\\
BD+21 4695	&	0.002	&	155	&	0.001	&	0.002	&	27	&	1	&	0.001	&	0.005	&	34	&	1	\\
BD+23 1148	&	0.032	&	164	&	0.032	&	0.002	&	35.26	&	0.01	&	0.029	&	0.005	&	41.04	&	0.02	\\
BD+25 4083	&	0.016	&	166	&	0.0101	&	0.0005	&	19.86	&	0.04	&	0.004	&	0.004	&	78	&	1	\\
BD+27 797	&	0.017	&	8	&	0.006	&	0.005	&	26.7	&	0.5	&	0.01	&	0.07	&	163	&	9	\\
BD+27 850	&	0.009	&	149	&	0.004	&	0.001	&	5.1	&	0.8	&	0.01	&	0.01	&	178	&	23	\\
BD+27 3411	&	0.010	&	153	&	0.007	&	0.001	&	7.9	&	0.4	&	0.00	&	0.09	&	107	&	32	\\
BD+28 3598	&	0.051	&	40	&	0.030	&	0.002	&	12.86	&	0.06	&	0.01	&	0.02	&	114	&	1	\\
BD+29 3842	&	0.017	&	80	&	0.0038	&	0.0002	&	3.9	&	0.2	&	0.00	&	0.01	&	26	&	5	\\
BD+29 4453	&	0.010	&	39	&	0.001	&	0.002	&	30.4	&	0.8	&	0.004	&	0.006	&	8	&	4	\\
BD+30 3227	&	0.016	&	65	&	0.0068	&	0.0005	&	21.78	&	0.05	&	0.00	&	0.01	&	125	&	1	\\
BD+31 4018	&	0.014	&	80	&	0.039	&	0.006	&	20.5	&	0.1	&	0.003	&	0.008	&	86	&	10	\\
BD+36 3946	&	0.007	&	168	&	0.052	&	0.004	&	20.11	&	0.06	&	0.021	&	0.006	&	81.8	&	0.5	\\
BD+37 675	&	0.005	&	147	&	0.12	&	0.01	&	4.6	&	0.3	&	0.12	&	0.01	&	94.4	&	0.4	\\
BD+37 3856	&	0.014	&	127	&	0.011	&	0.001	&	9.8	&	0.2	&	0.007	&	0.007	&	100	&	2	\\
BD+40 1213	&	0.015	&	165	&	0.008	&	0.003	&	17.2	&	0.3	&	0.00	&	0.01	&	114	&	2	\\
BD+42 1376	&	0.015	&	175	&	0.014	&	0.001	&	27.40	&	0.04	&	0.01	&	0.01	&	129.9	&	0.2	\\
BD+42 4538	&	0.008	&	86	&	0.0043	&	0.0002	&	5.2	&	0.1	&	0.003	&	0.004	&	83	&	3	\\
BD+43 1048	&	0.025	&	88	&	0.023	&	0.002	&	22.48	&	0.05	&	0.015	&	0.004	&	77.2	&	0.3	\\
BD+45 933	&	0.023	&	138	&	0.042	&	0.002	&	20.75	&	0.03	&	0.015	&	0.005	&	71.4	&	0.3	\\
BD+45 3879	&	0.006	&	24	&	0.0074	&	0.0006	&	3.6	&	0.3	&	0.007	&	0.004	&	68.7	&	0.4	\\
BD+46 275	&	0.009	&	89	&	0.0021	&	0.0001	&	4.8	&	0.2	&	0.027	&	0.007	&	10.4	&	0.4	\\
BD+47 183	&	0.011	&	82	&	0.0043	&	0.0003	&	36.82	&	0.01	&	0.01	&	0.04	&	158	&	5	\\
BD+47 857	&	0.002	&	64	&	0.0050	&	0.0006	&	29.39	&	0.05	&	0.01	&	0.06	&	170	&	10	\\
BD+47 939	&	0.011	&	174	&	0.007	&	0.001	&	29.36	&	0.07	&	0.004	&	0.004	&	54.0	&	0.2	\\
BD+47 3985	&	0.004	&	53	&	0.0039	&	0.0008	&	5.1	&	0.6	&	0.007	&	0.008	&	10	&	2	\\
BD+49 614	&	0.012	&	104	&	0.014	&	0.002	&	23.4	&	0.1	&	0.078	&	0.007	&	9.9	&	0.1	\\
BD+50 825	&	0.007	&	146	&	0.041	&	0.003	&	20.41	&	0.05	&	0.012	&	0.006	&	68.3	&	0.3	\\
BD+50 3430	&	0.005	&	10	&	0.005	&	0.001	&	13.5	&	0.2	&	0.004	&	0.004	&	79	&	1	\\
BD+51 3091	&	0.007	&	22	&	0.0020	&	0.0003	&	20.91	&	0.09	&	0.003	&	0.004	&	49.9	&	0.2	\\
BD+53 2599	&	0.006	&	33	&	0.023	&	0.002	&	17.83	&	0.07	&	0.010	&	0.006	&	11.7	&	0.9	\\
BD+55 552	&	0.027	&	112	&	0.009	&	0.001	&	36.84	&	0.03	&	0.004	&	0.006	&	144.5	&	0.3	\\
BD+55 605	&	0.042	&	121	&	0.0013	&	0.0002	&	28.49	&	0.06	&	0.031	&	0.006	&	3	&	1	\\
BD+55 2411	&	0.002	&	149	&	0.0009	&	0.0001	&	20.29	&	0.07	&	0.001	&	0.006	&	29	&	2	\\
BD+56 473	&	0.047	&	109	&	0.009	&	0.003	&	26.5	&	0.1	&	0.005	&	0.006	&	171	&	2	\\
BD+56 478	&	0.037	&	112	&	0.0032	&	0.0002	&	10.90	&	0.06	&	0.005	&	0.005	&	46.85	&	0.04	\\
BD+56 484	&	0.037	&	124	&	0.012	&	0.002	&	12.4	&	0.2	&	0.00	&	0.02	&	164	&	4	\\
BD+56 493	&	0.044	&	122	&	0.0024	&	0.0004	&	15.0	&	0.1	&	0.003	&	0.005	&	45.66	&	0.02	\\
BD+56 511	&	0.041	&	114	&	0.0026	&	0.0001	&	19.27	&	0.04	&	0.01	&	0.01	&	175	&	8	\\
BD+56 573	&	0.042	&	119	&	0.047	&	0.003	&	20.12	&	0.05	&	0.009	&	0.005	&	90.0	&	0.9	\\
BD+57 681	&	0.060	&	128	&	0.0106	&	0.0005	&	37.25	&	0.01	&	0.007	&	0.005	&	19.3	&	0.5	\\
BD+58 554	&	0.033	&	117	&	0.0040	&	0.0004	&	28.60	&	0.04	&	0.009	&	0.005	&	176	&	3	\\
BD+58 2320	&	0.017	&	60	&	0.016	&	0.001	&	25.66	&	0.03	&	0.038	&	0.004	&	45.843	&	0.002	\\
        \hline
	\end{tabular}
\end{table*}



\bsp	
\label{lastpage}
\end{document}